\documentclass[%
 reprint,
superscriptaddress,
nolongbibliography,
 amsmath,amssymb,
 aps,
floatfix,
]{revtex4-2}

\usepackage{color}
\usepackage{graphicx}
\usepackage{dcolumn}
\usepackage{bm}
\usepackage[colorlinks= true,urlcolor=blue,linkcolor= blue,citecolor=blue,bookmarks=false,pdfstartview=]{hyperref}
\usepackage{wasysym}
\usepackage{amstext}
\usepackage{gensymb}
\usepackage{amssymb}
\usepackage{soul}
\usepackage{color}

\begin{document}

\preprint{APS/123-QED}

\title{Emergence of a Fluctuating Ground State in Y-kapellasite under Pressure}
\author{Dipranjan Chatterjee}
 \email{dipranjan.chatterjee@physics.ox.ac.uk}
\affiliation{Universit\'{e}  Paris-Saclay,  CNRS,  Laboratoire  de  Physique  des  Solides,  91405,  Orsay,  France}
\affiliation{Clarendon Laboratory, Department of Physics, University of Oxford, Parks Road, OX1 3PU, Oxford, United Kingdom}
\author{Petr Doležal}
\affiliation{Department of Condensed Matter Physics, Faculty of Mathematics and Physics, Charles University, Ke Karlovu 5, 121 16 Prague 2, Czech Republic}
\affiliation{Institute of Solid State Physics, TU Wien, 1040 Vienna, Austria}
\author{Federico Abbruciati}
\affiliation{European Synchrotron Radiation Facility, 71 Avenue des Martyrs, F-38043 Grenoble, France}
\author{Tobias Biesner}
\affiliation{1.~Physikalisches Institut, Universit\"at Stuttgart, Pfaffenwaldring 57, 70569 Stuttgart, Germany}
\author{Katharina M. Zoch}
\affiliation{Physikalisches Institut, Goethe-Universit\"at Frankfurt, Frankfurt am Main, Germany}
\author{Rustem Khasanov}
\affiliation{Laboratory for Muon-Spin Spectroscopy, Paul Scherrer Institut, 5232 Villigen, Switzerland}
\author{Shams Sohel Islam}
\affiliation{Laboratory for Muon-Spin Spectroscopy, Paul Scherrer Institut, 5232 Villigen, Switzerland}
\author{Guratinder Kaur}
\affiliation{Max-Planck-Institute for Solid State Research, Heisenbergstra{\ss}e 1, 70569 Stuttgart, Germany}
\author{Seulki Roh}
\affiliation{1.~Physikalisches Institut, Universit\"at Stuttgart, Pfaffenwaldring 57, 70569 Stuttgart, Germany}
\author{Francesco Capitani}
\affiliation{Synchrotron SOLEIL, L’Orme des Merisiers D\'epartementale 128, 91192 Saint-Aubin, France}
\author{Joao Elias F. S. Rodrigues}
\affiliation{European Synchrotron Radiation Facility, 71 Avenue des Martyrs, F-38043 Grenoble, France}%
\author{Gaston Garbarino}
\affiliation{European Synchrotron Radiation Facility, 71 Avenue des Martyrs, F-38043 Grenoble, France}%
\author{Cornelius Krellner}
\affiliation{Physikalisches Institut, Goethe-Universit\"at Frankfurt, Frankfurt am Main, Germany}
\author{Philippe Mendels}
\affiliation{Universit\'{e}  Paris-Saclay,  CNRS,  Laboratoire  de  Physique  des  Solides,  91405,  Orsay,  France}%
\author{Edwin Kermarrec}
\affiliation{Universit\'{e}  Paris-Saclay,  CNRS,  Laboratoire  de  Physique  des  Solides,  91405,  Orsay,  France}%
\author{Martin Dressel}
\affiliation{1.~Physikalisches Institut, Universit\"at Stuttgart, Pfaffenwaldring 57, 70569 Stuttgart, Germany}%
\author{Bj\"orn Wehinger}
\affiliation{European Synchrotron Radiation Facility, 71 Avenue des Martyrs, F-38043 Grenoble, France}%
\author{Andrej Pustogow}
\affiliation{Institute of Solid State Physics, TU Wien, 1040 Vienna, Austria}%
\author{Fabrice Bert}
\affiliation{Universit\'{e}  Paris-Saclay,  CNRS,  Laboratoire  de  Physique  des  Solides,  91405,  Orsay,  France}%
\author{Pascal Puphal}
\email{P.Puphal@fkf.mpg.de}
\affiliation{Max-Planck-Institute for Solid State Research, Heisenbergstra{\ss}e 1, 70569 Stuttgart, Germany}
\affiliation{1.~Physikalisches Institut, Universit\"at Stuttgart, Pfaffenwaldring 57, 70569 Stuttgart, Germany}%

\date{\today}%

\begin{abstract}
  Y-kapellasite (Y$_3$Cu$_9$(OH)$_{19}$Cl$_8$), which hosts an original anisotropic kagome sublattice, is a promising candidate for studying elusive and complex correlated physics. It exhibits a theoretically predicted in-plane $(1/3, 1/3)$ magnetic order \cite{Hering2022}  but its magnetic interaction values place it close to a phase boundary to a spin liquid state~\cite{Chatterjee2023}. Our $\mu$SR measurements under hydrostatic pressure demonstrate the complete suppression of static magnetism in favor of a fully dynamical ground state at $2.3$~GPa. Complementary high-pressure x-ray and optical phonon measurements reveal a gradual reduction of the kagome anisotropy, enhancing magnetic frustration without structural transitions. Our results establish Y-kapellasite as a rare clean kagome model in which long-range order is suppressed by pressure-tuned frustration, the first fingerprint for the realization of a quantum spin liquid without strong disorder.

\end{abstract}
\maketitle

\footnotetext[1]{See Supplemental Material for further details.}

Geometric frustration is a key ingredient to prevent conventional magnetic ordering \cite{starykh2015unusual} and, in combination with enhanced fluctuations in low dimensional quantum magnets, to stabilize the long sought spin liquid ground states, which offer a platform for exploring long-range quantum entanglement and fractionalized excitations \cite{lacroix2011introduction,Balents10}. The search for quantum spin liquids (QSLs) in two-dimensional frustrated magnets has spurred extensive experimental efforts across a variety of geometries. Former work examined triangular systems such as $\kappa$-(BEDT-TTF)$_2$Cu$_2$(CN)$_3$ \cite{Shimizu2003} or YbMgGaO$_4$ \cite{Paddison2016,Li2016}, and  corner-sharing triangular kagome antiferromagnets including ZnCu$_3$(OH)$_6$Cl$_2$ \cite{Mendels2007,Han2016}, ZnCu$_3$(OH)$_6$FBr \cite{Wang2021}, and ZnCu$_3$(OH)$_6$FCl \cite{Feng2018}, all of which display experimental signatures of QSL behavior. In these systems, however, the determination of the true ground state is complicated by chemical or structural disorder, which can mimic the features expected for a perfectly frustrated QSL \cite{Pinteric2014,Dressel2016,Miksch2021,Komatsu1996,Pustogow2022,Kremer2025,Bordelon2019,Zhu2017}.

Among kagome antiferromagnets, herbertsmithite ZnCu$_3$(OH)$_6$Cl$_2$ stands as the archetypal example, extensively studied with a range of experimental probes~\cite{sherman2016nuclear,Khuntia2020,Wang2021,PhysRevX.12.011014,PhysRevB.107.054434}. Inspired by Anderson’s proposal of the resonating valence bond (RVB) state as a fundamental ground state for undoped cuprates \cite{anderson1987}, chemical substitution in herbertsmithite has been pursued to tune its quantum state. One notable derivative, Y-kapellasite Y$_3$Cu$_9$(OH)$_{19}$Cl$_8$, is an insulating kagome antiferromagnet in which a lattice distortion arises from the displacement of Y atoms, partially releasing the intrinsic frustration~\cite{Puphal2017}. As a result, the magnetic ions Cu$^{2+}$ (S=1/2) form an original anisotropic kagome lattice, modeled with three distinct nearest-neighbor interactions~\cite{Hering2022}. Detailed NMR, $\mu$SR, and neutron scattering studies~\cite{Chatterjee2023,Wang2023} have established in-plane $(1/3,\,1/3)$ magnetic order in a disorder-free single crystal. Further, the magnetic interactions, obtained by spin wave analysis, place the compound in the vicinity of the boundary to a spin liquid phase in the phase diagram computed classically, thus motivating the present high-pressure investigations \cite{Chatterjee2023}.

Pressure is a powerful nonthermal tuning parameter in quantum materials, capable of altering bandwidth and correlations \cite{MottAFM-FM,mott2}, Fermi-surface topology \cite{pan2015pressure,kang2015superconductivity,Pengpressure}, magnetism, Kondo physics, and phonon interactions \cite{heavyferm-1,knebel2008quantum,sakrikar2025pressure}. In frustrated magnets, it can be used to separate the roles of disorder and geometric frustration. For example, in pyrochlores, pressure has been shown to induce the coexistence of long-range order and QSL behavior in Tb$_2$Ti$_2$O$_7$ \cite{Mirebeau2002}, drive a transition from QSL to ferromagnetism in Yb$_2$Ti$_2$O$_7$ \cite{Kermarrec2017}, and suppress ferromagnetic correlations in Tb$_2$Sn$_2$O$_7$ \cite{Mirebeau2009}. Pressure can also reduce disorder, as in the Kitaev candidate material $\alpha$-RuCl$_3$ \cite{Stahl2024}. In herbertsmithite, the crystal symmetry and, hence, the perfect kagome lattice are preserved, with only a moderate change of the Cu-O-Cu bond angles, up to 8~GPa until a structural transition occurs~\cite{Malavi2020}. Accordingly, the $\mu$SR relaxation and susceptibility remain unchanged under pressure showing that the QSL ground state persists up to at least 3~GPa~\cite{Barthelemy2020,Malavi2020}.
In contrast, the present work examines a kagome antiferromagnet that is both geometrically frustrated and free from site disorder,  providing a rare opportunity to isolate and directly probe the role of pure geometric frustration in stabilizing a spin liquid ground state.

\begin{figure}[hbt!]
\begin{centering}
\includegraphics[width=0.8\columnwidth]{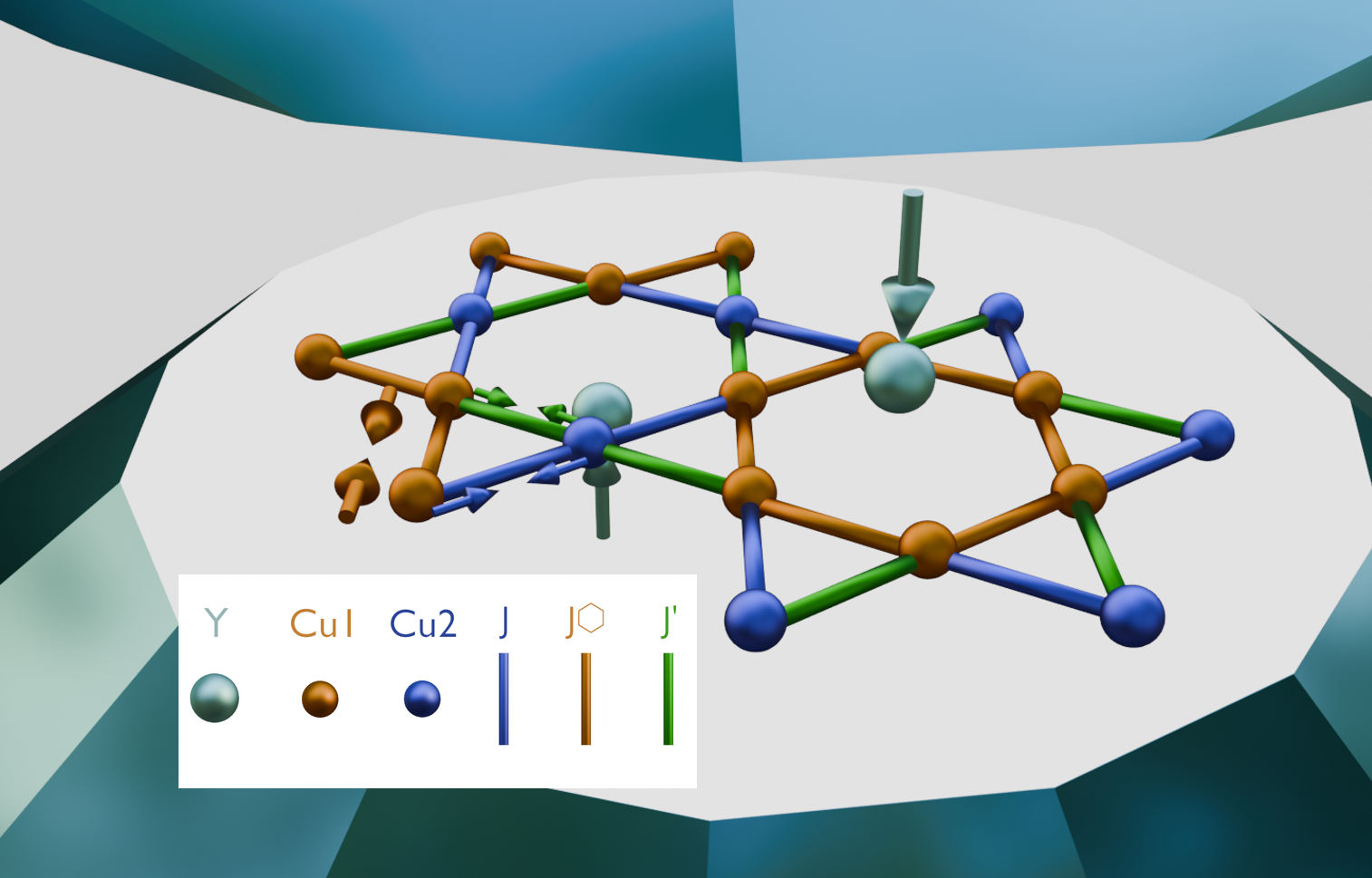}
\par\end{centering}
\caption{
Schematic view of a single kagome layer of Y-kapellasite inside the diamond anvil cell, highlighting the pressure-tuned magnetic ``skeleton'' made of two inequivalent Cu sites (Cu1 in orange, Cu2 in blue). The dominant exchange interaction \(J_{\hexagon}\) along Cu1--Cu1 bonds (orange), the weaker Cu1--Cu2 interaction \(J'\) (green), and the Cu2--Cu1 interaction \(J\) (blue) are indicated. Gray arrows illustrate the effect of hydrostatic pressure, which pushes the Y ions (located at the centers of the hexagons) back toward the kagome plane.  
}
\label{kag}
\end{figure}

Y$_3$Cu$_9$(OH)$_{19}$Cl$_8$, Y-kapellasite features a $1/9$ substitution of Cl$^-$ by (OH)$^-$, as compared to YCu$_3$(OH)$_{6}$Cl$_3$ and the shorter Y--OH bonds displace Y atoms out of the kagome plane, distorting the lattice into an anisotropic network with three distinct exchange interactions (Fig.~\ref{kag}). The resulting kagome ``skeleton,'' comprises two inequivalent Cu sites and anisotropic exchange interactions. The structural distortion shortens one Cu1--Cu2 bond to 3.26~\AA\ (green bond), while the remaining Cu--Cu distances remain near 3.38~\AA. Concomitantly, the Cu--O--Cu bond angle associated with this shortened bond is reduced to \(\sim110^\circ\), whereas the other angles remain close to \(116^\circ\)--\(117^\circ\) (Fig.~\ref{XRD}(c)). The magnetic sublattice can thus be viewed as Cu1--Cu1 hexagons (orange) coupled by three distinct interactions: \(J_{\hexagon}\) within the hexagons, \(J\) along Cu2--Cu1 bonds (blue), and a weaker \(J'\) along the shortened Cu1--Cu2 bonds (green). From our previous spin wave analysis \cite{Chatterjee2023}, we estimated \(J_{\hexagon}=J=-140(10)\)~K and \(J'=-63(7)\)~K, consistent with the hierarchy inferred from the bond angles. These interactions yield a Curie--Weiss temperature of \(-114(9)\)~K, in agreement with independent susceptibility measurements, and place Y-kapellasite in the \((1/3,1/3)\) antiferromagnetic phase, close to a boundary of a spin liquid regime \cite{Hering2022}.

 Thus, the system exhibits long-range order (LRO) below $T_N=2.2$~K~\cite{Chatterjee2023}. Uniaxial strain releases the  frustration and strengthens the antiferromagnetic order~\cite{Wang2023}. On the other hand, disorder can induce a spin liquid  state in powders~\cite{Barthelemy2019}. It is also present in the Br analog \cite{zeng2022possible,Liu2022,Zeng2024}, where a partial occupancy of the Y site occurs in the kagome plane, thus presenting a mixed system between YCu$_3$(OH)$_6$Cl$_3$ (in-plane Y) and Y$_3$Cu$_9$(OH)$_{19}$Cl$_8$ (out of plane Y). In contrast, our single crystals are free from site mixing, maintain a stable Cu$^{2+}$ oxidation state, and possess a well-defined superstructure, resulting in LRO confirmed by NMR, INS, and $\mu$SR \cite{Chatterjee2023,Wang2023} (see Supplemental Material for additional details \cite{Note1}).  Here we study these clean crystals under hydrostatic pressure, using $\mu$SR and complementary techniques to investigate how the evolution of the structure enhances the magnetic frustration and drives the nature of the ground state, in stark contrast to chemical pressure resulting in reduced frustration \cite{Krieger2025} (see Supplemental Materials for more details \cite{Note1}).

Previous $\mu$SR measurements at ambient pressure \cite{Chatterjee2023} on single crystal of Y$_3$Cu$_9$(OH)$_{19}$Cl$_8$ revealed a bulk magnetic transition at $T_N$ with a broadly distributed and unusually small ($\sim$8.6~mT) internal field at base temperature. To probe the stability of this ordered state, we performed a pressure-dependent  $\mu$SR experiment on crushed single crystals in a hydrostatic piston cylinder cell on the GPD spectrometer at the Paul Scherrer Institute. In this setup, a fraction $f$ (50--70\%) of muons stop in the pressure-cell walls, producing a known background relaxation $P_{cell}(t)$ represented by the dashed line in Fig.~\ref{muSR}(a) that can be separated from the sample signal~\cite{khasanov2016high}.

\begin{figure*}
\begin{centering}
\includegraphics[width=2\columnwidth]{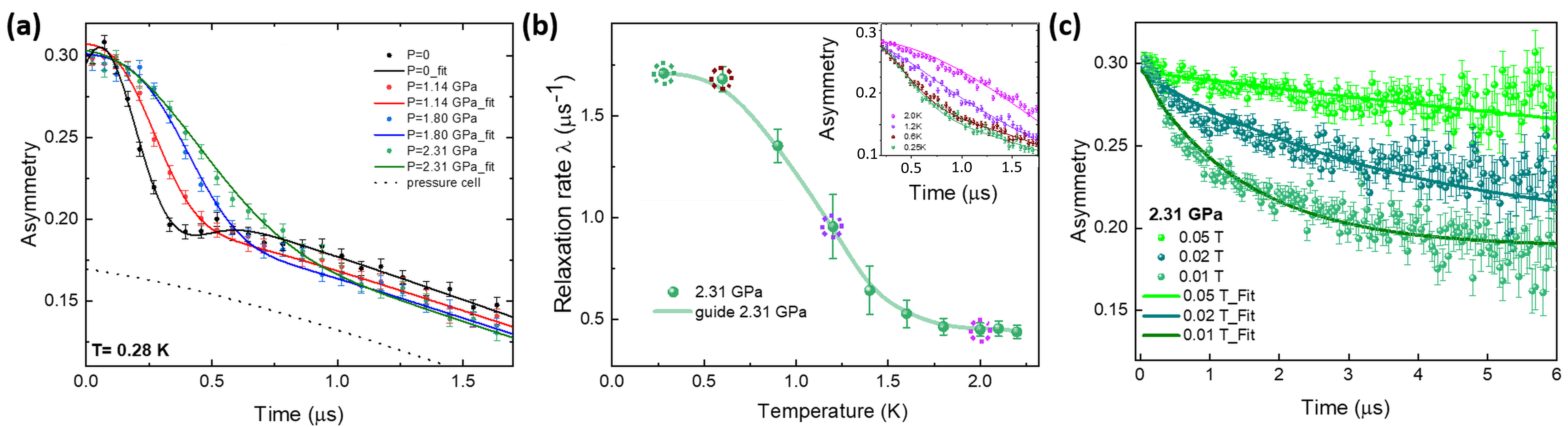}
\par\end{centering}
\caption{(a) Zero-field (ZF) asymmetries measured at $0.28\,\mathrm{K}$ under different pressures: ambient pressure, $1.14\,\mathrm{GPa}$, $1.8\,\mathrm{GPa}$, and $2.3\,\mathrm{GPa}$. The black dotted line indicates the contribution from the pressure cell. 
(b) Temperature evolution of the relaxation rate $\lambda$ at $P=2.3$ GPa. Inset: zero-field asymmetry spectra measured at $2.0\,\mathrm{K}$, $1.2\,\mathrm{K}$, $0.6\,\mathrm{K}$, and $0.25\,\mathrm{K}$ at $P=2.3$ GPa. (c) Time evolution of LF asymmetries at 0.28~K under three different fields: 0.01 T, 0.02 T, and 0.05 T at 2.3 GPa, along with the corresponding fits to the DKT model [see Eq.~\ref{eq:pressure_DKT}] demonstrating a dynamical ground state.
 }
\label{muSR}
\end{figure*}

Figure~\ref{muSR}(a) presents the zero-field (ZF) $\mu$SR spectra measured at the base temperature $T = 0.28$~K under ambient pressure and applied pressures of $1.14$, $1.8$, and $2.3$~GPa; no hysteresis is observed (see Supplemental Materials for more details \cite{Note1}), as the ambient pressure spectrum is fully recovered after releasing the applied pressure. As expected from previous measurements~\cite{Chatterjee2023}, the ZF spectrum at ambient pressure shows a pronounced dip near 0.4~$\mu$s, characteristic of static magnetic order. The cell fraction $f=0.53$ of the signal with the relaxation  $P_{\text{cell}}(t)$ was then determined by fitting the ambient pressure sample relaxation as in Ref.~\onlinecite{Chatterjee2023}, with a damped oscillatory term, $\frac{2}{3} \cos(\omega_f t + \phi) e^{-\sigma^2 t^2/2} + \frac{e^{-\lambda_f t}}{3}$, where $\omega_f$ is the muon precession frequency proportional to the internal field, $\sigma_f$ the field distribution width, and $\lambda_f$ the relaxation rate of the $1/3$ tail signal. 
Remarkably, at the maximum applied pressure, the dip disappears, and the asymmetry relaxation is markedly slower, indicating the suppression of static magnetism and the emergence of a dynamical ground state. 
We further measured the temperature dependence of the ZF asymmetries at $P=2.3$~GPa, which we could model as  
\begin{equation}
a(t) = A_0 \left[ f P_{\text{cell}}(t) + (1-f)\left(f_r e^{-(\lambda t)^\alpha} + (1-f_r)\right) \right],
\label{eq:pressure_ZF}
\end{equation}  
where $A_0 \sim 0.3$ is the initial asymmetry, $f = 0.54(2)$ the cell fraction, and $(1-f_r)=0.25$ a minority of weakly relaxing muons \cite{Barthelemy2019,Chatterjee2023}. For $T \gtrsim 1.6$~K, rapid Cu$^{2+}$ spin fluctuations cause motional narrowing, leaving $\lambda$ dominated by weak nuclear fields and nearly temperature independent as shown in Fig.~\ref{muSR}(b). At lower temperature, $\lambda(T)$ evolves smoothly into a dynamical plateau, as observed for all QSL candidates~\cite{Mendels2007,orain2014musr,faak2012kapellasite} and is taken as an indication of the persistent dynamics of correlated spins. 

To confirm that the low temperature spin correlations are fully dynamic at 2.3~GPa, we performed additional longitudinal-field (LF) $\mu$SR experiment (Fig.~\ref{muSR}(c)). Fields $H_{LF} > 0.01$~T fully decouple the pressure-cell background~\cite{khasanov2016high} and the quasistatic nuclear contribution, leaving only the electronic spin relaxation. If the weak $\lambda = 1.7~\mu\text{s}^{-1}$ relaxation observed at low $T$ were of static origin, it should be fully suppressed by fields $\gtrsim 0.02$ T. This is clearly not the case, which demonstrates that the relaxation is of dynamical origin. Accordingly, the LF data shown in Fig.~\ref{muSR}(c) were modeled with the standard dynamical Kubo--Toyabe (DKT) function,
\begin{equation}
a_{LF}(t) = (A_0-A_{bkg})P_{DKT}(t,\nu,\Delta,H_{LF}) + A_{bkg},
\label{eq:pressure_DKT}
\end{equation}
where $A_{bkg}\sim A_0f$, $\Delta/\gamma_\mu$ is the width of the distribution of the internal fields and $\nu$ their fluctuation rate~\cite{Hayano79}. Fits of the 0.01, 0.02 and 0.05~T data give $\Delta/\gamma_\mu = 2.9$~mT ($\Delta=2.41(3)$~MHz) and $\nu=7.81(2)$~MHz. Since $\nu>\Delta$, the 2.3~GPa state lies deep in the fluctuating regime, with no evidence of spin freezing. 
The absence of static magnetism despite strong exchange interactions provides direct evidence for a fluctuating, spin-liquid-like, ground state at 2.3~GPa in Y-kapellasite. We note that this ground state differs from the one of powder samples formerly studied in Ref.~\cite{Barthelemy2019}. Indeed, similar decoupling experiments revealed an inhomogeneous behavior, demonstrating residual disorder in powder samples, at variance with the homogeneous fluctuating ground state induced by pressure in crystals.

To elucidate the pressure-induced dynamical ground state and assess the role of geometric frustration, we investigated the kagome lattice of Y-kapellasite under external pressure using high-pressure single crystal x-ray diffraction (XRD). At ambient pressure, Y-kapellasite crystallizes in the trigonal space group No.~148, \(R\overline{3}\), with lattice parameters \(a=b=11.57\)~\AA\ and \(c=17.33\)~\AA. This structure constitutes a supercell and subgroup of YCu$_3$(OH)$_6$Cl$_3$ \cite{Kremer2025}. The zonal (0kl) diffraction maps of the subcell differ markedly from those of the supercell, and remain, independent of pressure as visible in Supplemental Material \cite{Note1}. Thus upon application of hydrostatic pressure, the kagome lattice remains isostructural, both at room temperature and at 3~K. However, subtle yet systematic changes in bond lengths and angles modify the magnetic superexchange interactions in an anisotropic fashion. In general, softer bonds are expected to compress more strongly, while bonds already under compression remain relatively rigid. Consistent with this expectation, the short Cu1-Cu2 bond (green) exhibits only a weak pressure dependence, whereas the longer Cu1-Cu1 bond (orange) contracts significantly under pressure [Fig.~\ref{XRD}(b)]. Similarly, the Cu1--O--Cu1 bond angle decreases substantially, while the Cu1-O-Cu2 angle remains essentially unchanged [Fig.~\ref{XRD}(c)]. These trends reflect the pressure-induced relaxation of the Y ions back toward the kagome plane, predominantly affecting the Cu1 sites associated with the dominant interaction \(J_{\hexagon}\).

\begin{figure}[hbt!]
\begin{centering}
\includegraphics[width=1\columnwidth]{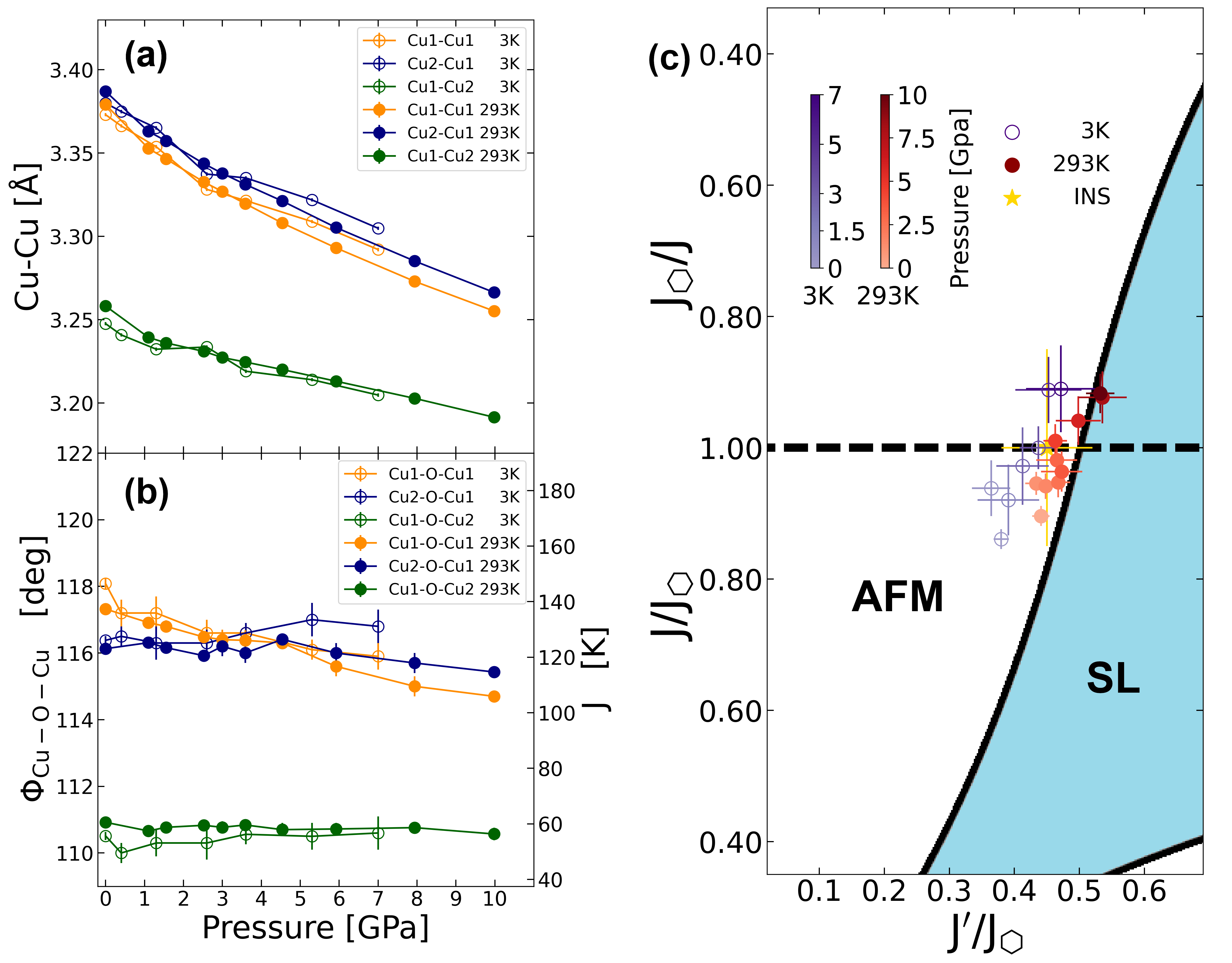}
\par\end{centering}
\caption{
(a) Pressure dependence of the refined Cu-Cu bond lengths: Cu1--Cu1 (orange), Cu1-Cu2 (green), and Cu2-Cu1 (dark blue).  
(b) Corresponding pressure evolution of the Cu-O-Cu bond angles defining the superexchange pathways \(J_{\hexagon}\) (Cu1-O-Cu1, orange), \(J'\) (Cu1-O-Cu2, green), and \(J\) (Cu2--O--Cu1, dark blue).  
(c) Classical phase diagram of the distorted kagome antiferromagnet as a function of the exchange interactions \(J_{\hexagon}\), \(J\), and \(J'\) \cite{Hering2022}. The white region denotes the \((1/3,1/3)\) antiferromagnetic ordered phase, while the blue region corresponds to a spin liquid (SL) regime. The exchange parameters obtained from spin wave analysis of inelastic neutron scattering data are shown as a yellow star \cite{Chatterjee2023}. Estimates of the pressure-dependent exchange couplings derived from Eq.~(\ref{eq:AF}) are indicated by blue (low temperature) and red (high temperature) symbols, with the applied pressure color coded as indicated in the legend.
}

\label{XRD}
\end{figure}

In cuprate oxides with a single Cu oxidation state, magnetic exchange interactions are governed primarily by superexchange, which depends sensitively on the Cu-O-Cu bond angle \(\varphi\). In the present case, the exchange can be estimated as
\begin{equation}
J~(\mathrm{K}) \approx -1270.5 + 12\,\varphi ,
\label{eq:AF}
\end{equation}
as discussed further in the Supplemental Material \cite{Note1}. Using this empirical relation together with the measured pressure evolution of the bond angles, we trace the trajectory of Y-kapellasite within the magnetic phase diagram [Fig.~\ref{XRD}(d)]. Increasing pressure enhances the degree of geometric frustration, driving the system toward the spin liquid regime. While the moderate pressures accessed here (\(\sim2\)~GPa) do not yet place the material fully inside the spin liquid region inferred from $\mu$SR measurements, it is important to note that the phase boundary shown in Fig.~\ref{XRD}(d) is based on a classical analysis. Its precise location for a quantum \(S=1/2\) kagome antiferromagnet such as Y-kapellasite remains an open theoretical question.

Pressure tuning provides a natural framework for understanding the emergence of a dynamical state as the anisotropic kagome lattice evolves toward a more isotropic configuration. This highlights the structural integrity of the kagome lattice as a key controlling factor. Y-kapellasite is known to undergo a structural transition near 33~K, clearly manifested in thermal expansion, specific heat, and NMR measurements \cite{Chatterjee2023}, but not detected by x-ray diffraction, as the Cu-O framework (``skeleton'') remains intact and only subtle hydrogen rearrangements occur. Optical spectroscopy, by contrast, is directly sensitive to OH-related phonons. Indeed, Ref.~\onlinecite{Dolezal2024} reports the emergence of additional weak modes below 32~K in the spectral range dominated by hydrogen motion, for example between 3200 and 3500~cm$^{-1}$.

Motivated by these findings, we measured infrared absorbance spectra under quasihydrostatic pressure, both below and above the transition, in two temperature regimes (14--18~K and 100~K), as shown in Fig.~\ref{optics}. The absence of any abrupt change in the pressure dependence at fixed temperature in either regime indicates that the structural modification is triggered by cooling below 33~K and persists independently of applied pressure, up to at least 10~GPa. Under hydrostatic pressure, we observe both redshifts and blueshifts of phonon modes. Such behavior is expected in a low-symmetry crystal with many unconstrained structural parameters, which need not exhibit a simple correlation with volume changes (see dotted lines in Fig.~\ref{optics}).

Following Ref.~\cite{Dolezal2024}, we classify the phonon spectra into regions associated with different atomic motions, indicated by black arrows at the bottom of Fig.~\ref{optics}, with their pressure evolution traced by gray dashed lines. Region~2 is dominated by Cu and O vibrations, whereas regions~3 and~4 primarily involve H and O motion. While both redshifts and blueshifts are readily observed in regions~3 and~4, the phonon modes in region~2 exhibit exclusively blueshifts. This reflects the rigidity of the Cu--O bonds forming the structural backbone of the crystal and stabilizing the kagome lattice of Cu$^{2+}$ ions.

In contrast, hydrogen atoms are bonded to a single oxygen and are therefore less constrained, consistent with the coexistence of redshifting and blueshifting modes. A representative example is region~4, which contains six calculated phonon modes \cite{Dolezal2024}. These modes begin to separate under pressure already at room temperature. Upon cooling to 14~K, additional modes emerge, which further split and sharpen with increasing pressure. Importantly, applying pressure at low temperature does not induce additional splittings or suppress any phonon modes. In regions~3 and~4, no unresolved hydrogen-related transitions remain. We, thus, conclude that the Cu--O framework remains essentially intact, aside from its continuous pressure-induced evolution observed by x-ray diffraction.

\begin{figure}
\begin{centering}
\includegraphics[width=1\columnwidth]{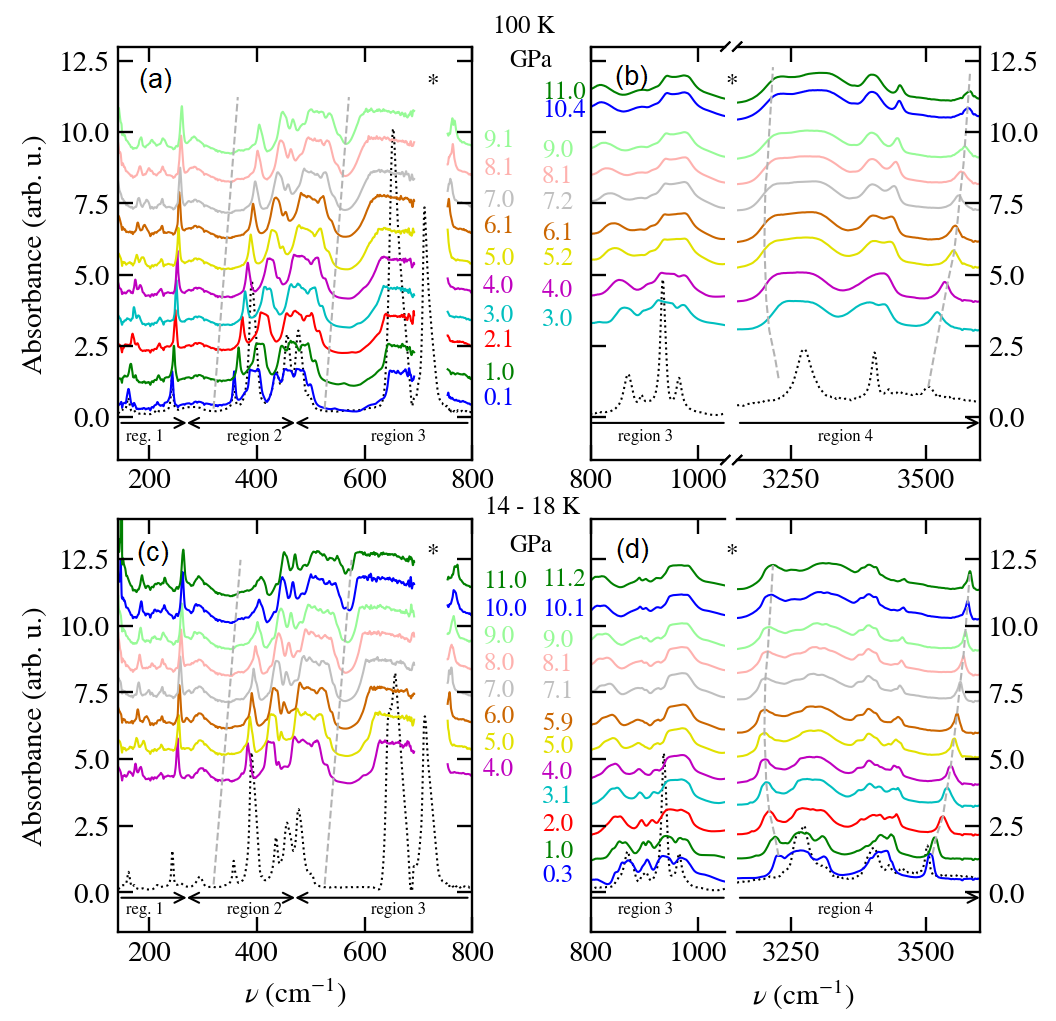}
\par\end{centering}
\caption{Low temperature absorbance with increasing pressure at 100~K (a),(b) and at 14-18~K (c),(d). The spectra are shifted along the $y$-axis for clarity. The regions marked by * are excluded because of the contribution from the experimental setup. The dotted line corresponds to the ambient pressure results at 100~K and 20~K, respectively. Far (1,2) and midinfrared (3,4) regions are indicated by horizontal double arrows separating the regions with the dominant motion of different atoms in the phonon modes. The gray dashed lines are guides for the eyes to highlight the blueshifts and redshifts.
}
\label{optics}
\end{figure}

To conclude, we realize the first controlled suppression of magnetic order into a spin-liquid-like, dynamical state, which we investigated by a comprehensive combination of x-ray, optical spectroscopy and $\mu$SR measurements under hydrostatic pressure. The pressure-induced fluctuating ground state observed here in crystals should be contrasted with the inhomogeneous spin dynamics observed formerly in powder samples~\cite{Barthelemy2019}, where the spin-liquid-like ground state seems then to be related to some residual, uncontrolled, disorder. Moreover, chemical pressure enhances structural distortions and drives the system away from the SL regime, underscoring external pressure as a uniquely clean and precise tuning parameter. Thus our work establishes Y-kapellasite as an archetypal platform to study the rise and fall of frustrated magnetism in a clean system without disorder. Exploring this material in more detail and performing similar experimental endeavors on other frustrated materials constitutes a major step forward toward the realization of the long sought quantum spin liquid. This work paves the way to further experiments under pressure to give evidence for long range entanglement and fractionalization to characterize the nature of the fluctuating ground state and the possible realization of a quantum spin liquid. Future work can shed light on its relation to the proposed spin liquid in related disordered Br analog compounds at ambient pressure.

\maketitle

\section*{Acknowledgments}
The XRD data sets can openly be accessed at https://doi.esrf.fr/10.15151/ESRF-DC-2191835598. The crystallographic structures are available from the joint CCDC/FIZ Karlsruhe data base with deposition number 2467984-2468000. We acknowledge the ESRF within the in-house proposals IH-HC-3746 and IH-HC-4021 and SOLEIL for provision of synchrotron radiation. We acknowledge support by the Deutsche Forschungsgemeinschaft (DFG) via DR228/68-1 and TRR288 (No. 422213477, Project A03).  A.P. acknowledges support by Hochschuljubiläumsfonds der Stadt Wien (Grant No. H-918729/2022). We acknowledge the support of the French Agence Nationale de la Recherche, under Grant No. ANR-25-CE30-2010-01 “ULTIMAT” and the
Fondation Charles Defforey-Institut de France. We thank Jeroen Jacobs for preparation and gas loading of the high-pressure cells,  Apostolos  Pantousas for single crystal refinement support and Jens Jakschik for measurement support at the SOLEIL beamtime.The work of
P. D. was funded by the Czech Science Foundation (research Project No. GAČR 23-06810O). D.C. acknowledges funding from UK Research and Innovation (UKRI) through the UK government’s Horizon Europe funding guarantee (Grant No. EP/X025861/1) and thanks Prof. Stephen J. Blundell for his support.

\bibliography{library.bib}

\end{document}


\preprint{APS/123-QED}

\title{Suppplemental Material to: Emergence of a Spin Liquid Ground State in Y-Kapellasite under Pressure}
\author{Dipranjan Chatterjee}
 \email{dipranjan.chatterjee@physics.ox.ac.uk}
\affiliation{Universit\'{e}  Paris-Saclay,  CNRS,  Laboratoire  de  Physique  des  Solides,  91405,  Orsay,  France}
\affiliation{Clarendon Laboratory, Department of Physics, University of Oxford, Parks Road, OX1 3PU, Oxford, United Kingdom}
\author{Petr Doležal}
\affiliation{Department of Condensed Matter Physics, Faculty of Mathematics and Physics, Charles University, Ke Karlovu 5, 121 16 Prague 2, Czech Republic}
\affiliation{Institute of Solid State Physics, TU Wien, 1040 Vienna, Austria}
\author{Federico Abbruciati}
\affiliation{European Synchrotron Radiation Facility, 71 Avenue des Martyrs, F-38043 Grenoble, France}
\author{Tobias Biesner}
\affiliation{1.~Physikalisches Institut, Universit\"at Stuttgart, Pfaffenwaldring 57, 70569 Stuttgart, Germany}
\author{Katharina M. Zoch}
\affiliation{Physikalisches Institut, Goethe-Universit\"at Frankfurt, Frankfurt am Main, Germany}
\author{Rustem Khasanov}
\affiliation{Laboratory for Muon-Spin Spectroscopy, Paul Scherrer Institut, 5232 Villigen, Switzerland}
\author{Shams Sohel Islam}
\affiliation{Laboratory for Muon-Spin Spectroscopy, Paul Scherrer Institut, 5232 Villigen, Switzerland}
\author{Guratinder Kaur}
\affiliation{Max-Planck-Institute for Solid State Research, Heisenbergstra{\ss}e 1, 70569 Stuttgart, Germany}
\author{Seulki Roh}
\affiliation{1.~Physikalisches Institut, Universit\"at Stuttgart, Pfaffenwaldring 57, 70569 Stuttgart, Germany}
\author{Francesco Capitani}
\affiliation{Synchrotron SOLEIL, L’Orme des Merisiers D\'epartementale 128, 91192 Saint-Aubin, France}
\author{Joao Elias F. S. Rodrigues}
\affiliation{European Synchrotron Radiation Facility, 71 Avenue des Martyrs, F-38043 Grenoble, France}%
\author{Gaston Garbarino}
\affiliation{European Synchrotron Radiation Facility, 71 Avenue des Martyrs, F-38043 Grenoble, France}%
\author{Cornelius Krellner}
\affiliation{Physikalisches Institut, Goethe-Universit\"at Frankfurt, Frankfurt am Main, Germany}
\author{Philippe Mendels}
\affiliation{Universit\'{e}  Paris-Saclay,  CNRS,  Laboratoire  de  Physique  des  Solides,  91405,  Orsay,  France}%
\author{Edwin Kermarrec}
\affiliation{Universit\'{e}  Paris-Saclay,  CNRS,  Laboratoire  de  Physique  des  Solides,  91405,  Orsay,  France}%
\author{Martin Dressel}
\affiliation{1.~Physikalisches Institut, Universit\"at Stuttgart, Pfaffenwaldring 57, 70569 Stuttgart, Germany}%
\author{Bj\"orn Wehinger}
\affiliation{European Synchrotron Radiation Facility, 71 Avenue des Martyrs, F-38043 Grenoble, France}%
\author{Andrej Pustogow}
\affiliation{Institute of Solid State Physics, TU Wien, 1040 Vienna, Austria}%
\author{Fabrice Bert}
\affiliation{Universit\'{e}  Paris-Saclay,  CNRS,  Laboratoire  de  Physique  des  Solides,  91405,  Orsay,  France}%
\author{Pascal Puphal}
\email{P.Puphal@fkf.mpg.de}
\affiliation{Max-Planck-Institute for Solid State Research, Heisenbergstra{\ss}e 1, 70569 Stuttgart, Germany}
\affiliation{1.~Physikalisches Institut, Universit\"at Stuttgart, Pfaffenwaldring 57, 70569 Stuttgart, Germany}%

\date{\today}%
\maketitle
\section{Supplemental Material}

The key experiments are based on pressure and temperature dependant  $\mu$SR, XRD and absorbance measurements. For visibility a pressure-Temperature phase diagram is created marking the individiual measurements in Fig. \ref{pT}. In the following the experimental methods and individual experimental details are discussed.

\setcounter{figure}{0}
\renewcommand{\figurename}{Fig.}
\renewcommand{\thefigure}{S\arabic{figure}}

\begin{figure}[h]
	\begin{centering}
		\includegraphics[width=1\columnwidth]{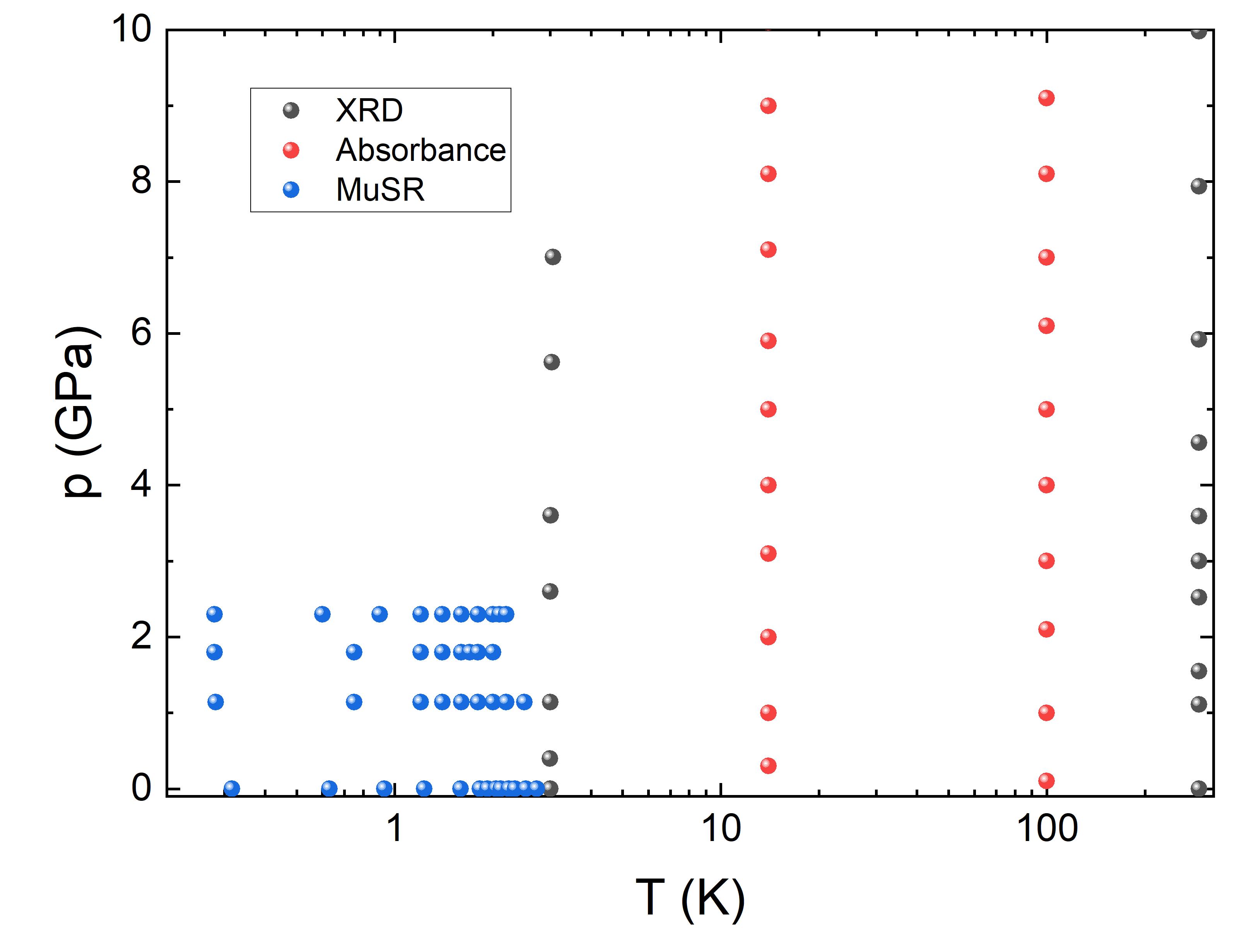} 
		\par\end{centering}
	\caption{p-T Experimental datapoints marked at various pressures versus a logarithmic scale of temperatures.}
	\label{pT}
\end{figure}

\subsection*{Synthesis}
Single-crystal growth is realized by slowly dissolving CuO in a YCl$_3$-H$_2$O solution and transporting it to the cold end. For this 4 g YCl$_3$-6H$_2$O, 2.3 g of CuO (Alfa Aesar, 99.9\%) and  5.5 ml H$_2$O were transferred in a quartz ampule with outer diameter of 30 mm and an inner one of 24 mm. The growth is executed in a three zone furnace with a gradient of 30$\degree$C and a temperature of 240$\degree$C at the hot end (note that the elevated pressure at this temperature requires especially thick quartz ampules). The gradient was optimized as too low temperatures yielded a mixture of Y-Kapellasite and Clinoatacamite. After 7 months the inclusion-free hexagonal single-crystals have an average size of 3x3x1~mm$^3$ up to 4x4x2~mm$^3$.

\subsection*{Muons}

\begin{figure*}[h]
	\begin{centering}
		\includegraphics[width=2\columnwidth]{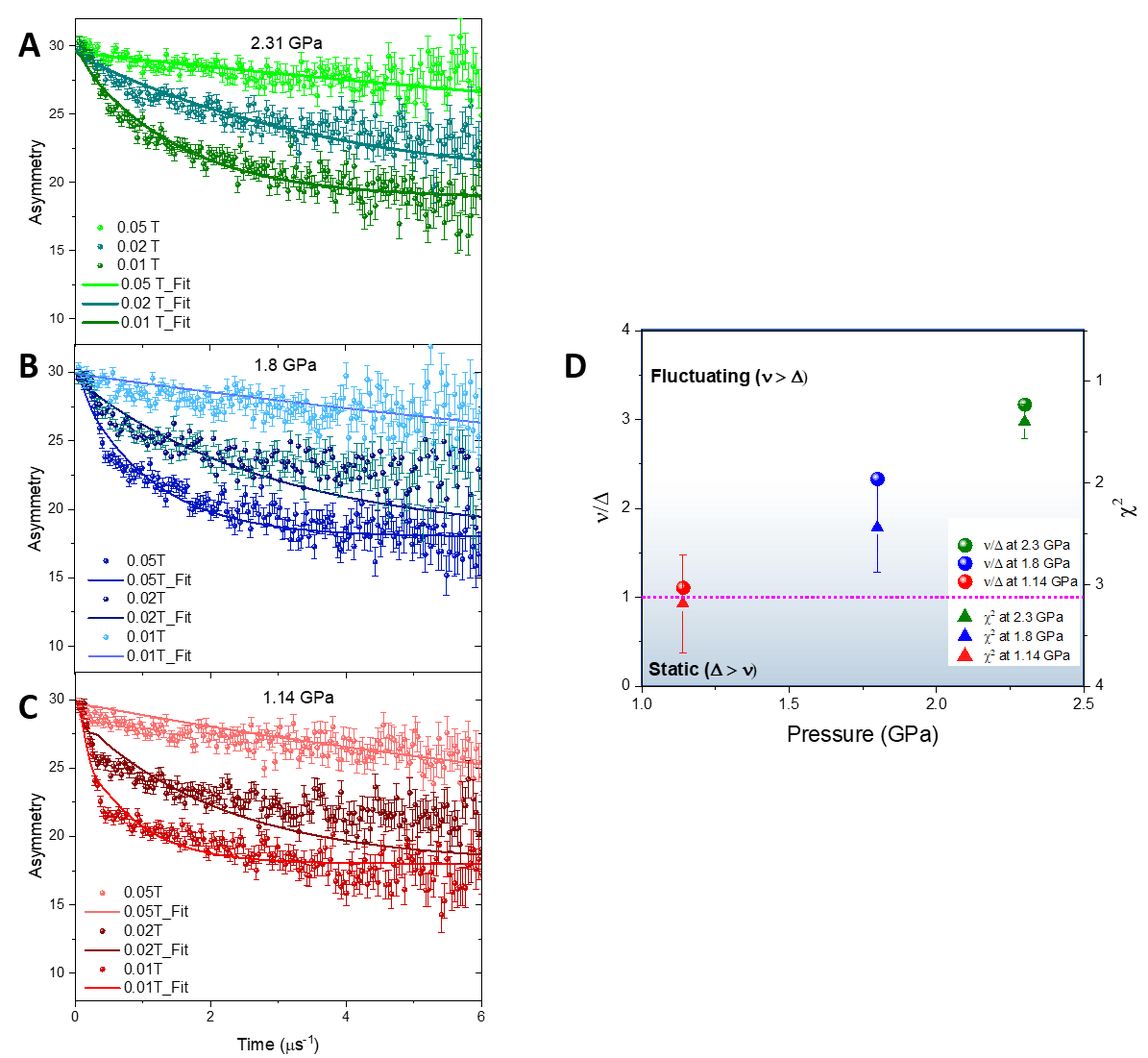}
		\par\end{centering}
	\caption{Assessment of pressure-induced enhanced fluctuations using the DKT model. Asymmetry with longitudinal fields of 0.05~T, 0.02~T, and 0.01~T fitted with the DKT model at 0.28~K and at pressures: (a) 2.3~GPa, (b) 1.8~GPa, and (c) 1.14~GPa. (d) The left $y$-axis shows the evolution of the dynamic ground state with pressure, quantified by the ratio $\nu/\Delta$, where $\nu$ represents the fluctuation rate of a field distribution of width $\Delta/\gamma_\mu$. The dotted violet line at $\nu/\Delta = 1$ marks the transition in the DKT model between dynamic and static regimes. The right $y$-axis depicts the deviation of the fits, represented by the $\chi^2$ values obtained from global fits in the left panels \textbf{a}, \textbf{b} and \textbf{c} . As pressure decreases, the fit quality deteriorates. This degradation arises as the system loses dynamicity, and the Dynamic Kubo-Toyabe model fails to capture the complex relaxation (different from a simple Kubo-Toyabe) in the quasi-static ground state at lower pressures.
 }
	\label{sup-mu}
\end{figure*}

ZF, TF and LF MuSR experiments were carried out at the Paul Scherrer Institute PSI, Switzerland. Experiments under the quasi-hydrostatic pressure conditions were conducted at the MuE1 beam-line using the GPD spectrometer \cite{khasanov2016high,Khasanov2022}. A pressure up to 2.3 GPa was generated in a double-walled clamp type cell made of nonmagnetic MP35N alloy. As a pressure transmitting medium, Daphne 7373 oil was used. We filled the pressure cell with about 1.7~g of crushed single-crystals for the measurements. The applied pressure was determined at low temperature by measuring the superconducting transition temperature of a small piece of In placed inside the cell. Daphne 7373 oil solidifies at around  $\simeq 2.3$~GPa at room temperature. This implies that certain effects due to nonhydrostaticity may be present at the highest measured pressure. From the broadening of the superconducting transition of elemental indium, we estimate a pressure inhomogeneity of roughly 0.7~kbar at the highest pressure studied.

 Having successfully fitted the asymmetry at 2.3 GPa for longitudinal fields of 0.01 T, 0.02 T, and 0.05 T using the dynamical Kubo-Toyabe (DKT) model, which clearly indicates a spin liquid ground state, we extended the application of the DKT model to intermediate pressures of 1.8 GPa and 1.14 GPa. The presence of mixed frozen and dynamic phases in the intermediate pressure regime results in a gradual degradation of the fits as the pressure decreases, as shown in figure \ref{sup-mu} \textbf{ a,b,c}. Specifically, the fluctuation rate \( \nu \) slows, while the static internal field strength, proportional to the parameter \( \Delta \), increases, as shown in Fig.~\ref{sup-mu} \textbf{d}. These trends continue as the system approaches ambient pressure, culminating in an ordered state.

\begin{figure}[h]
	\begin{centering}
		\includegraphics[width=1\columnwidth]{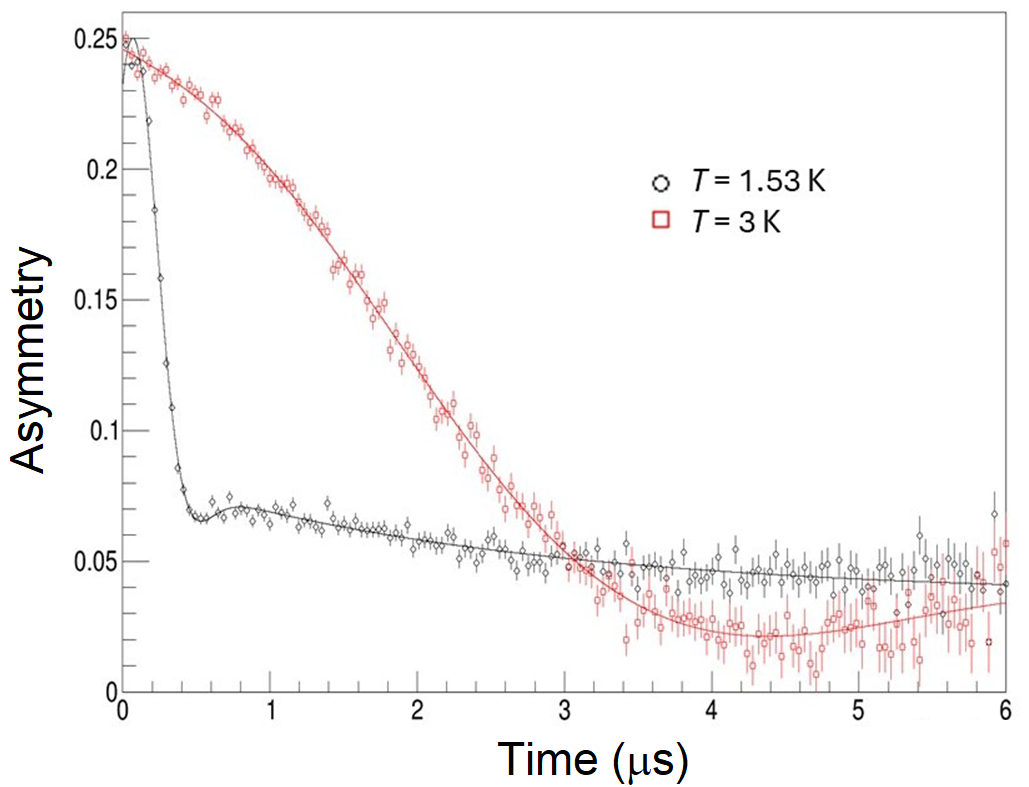}
		\par\end{centering}
	\caption{Post pressure experiment in ZF. Shown is the Asymmetry against the time. 
 }
	\label{sup-mu2}
\end{figure}

The pressure sequence in our experiment was 0 GPa → 2.3 GPa → 1.14 GPa → 1.8 GPa. Since the intermediate-pressure measurements were conducted after reaching the maximum pressure, and the relaxation rate and asymmetry nearly recovered at 1.14 GPa, this already suggested minimal hysteresis. In addition, we have after completing the pressure experiments verified the reversibility and performed zero-field MuSR measurements on the same crushed crystals sample on the GPS spectrometer at PSI (Fig.~\ref{sup-mu2}). The data show a full recovery of the rapidly damped oscillations observed in Y-kapellasite below its magnetic transition \cite{Chatterjee2023}.

\begin{figure}[h]
	\begin{centering}
        \includegraphics[width=1\columnwidth]{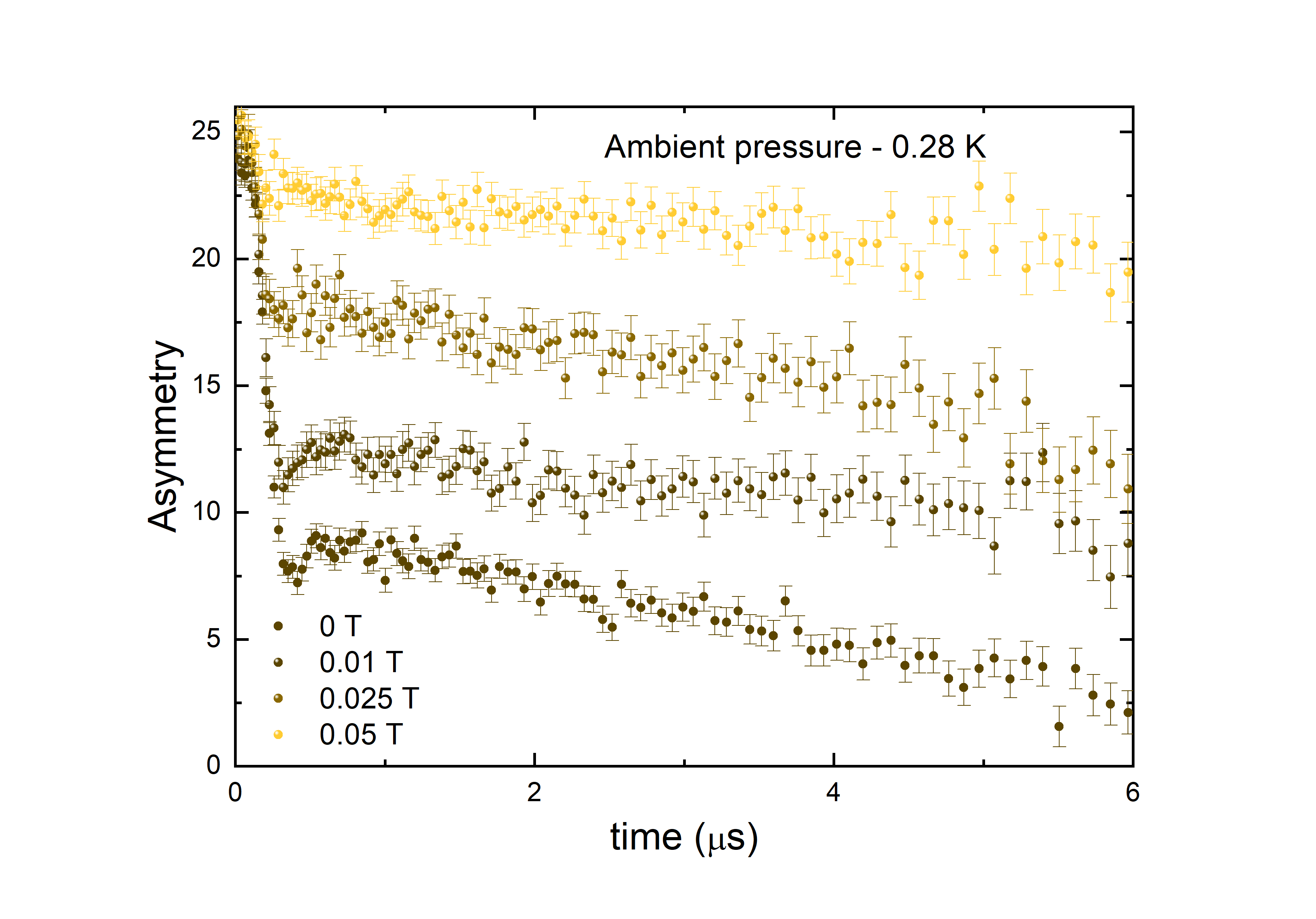}
		\par\end{centering}
    \caption{Zero and longitudinal-field $\mu$SR asymmetry at 0.28~K measured with the Dolly spectrometer at the Paul Scherrer Institute. Here, there is no pressure cell, the sample is measured in veto mode as in ref.~\cite{Chatterjee2023}.
 }
	\label{sup-mu3}
\end{figure}

For completeness we show on figure~\ref{sup-mu3} the decoupling experiment at ambient pressure in the ordered phase at 0.28~K. We recall that the 1/3rd tail residual relaxation is $\sim 0.05 \mu$s$^{-1}$~\cite{Chatterjee2023} much lower than the relaxation observed at 0.28~K at 2.3~GPa.

\subsection*{Diffraction}
Single crystal XRD data under applied pressure at all temperatures were collected at the beamline ID27 \cite{mezouar2024high} and the low temperature data at ambient pressure at ID15B \cite{ garbarino2024extreme} of the European Synchrotron Radiation Facility (ESRF).
We used membrane driven diamond anvil cells and helium as the pressure transmitting medium. Natural single crystalline diamonds with Boehler-Almax cut culets of 500 $\mu$m in size were mounted on seats with a modified Boehler-Almax design and an opening of 70$^\circ$. Stainless steel was used as gasket material. 200 $\mu$m thick plates were pre-indented to 100 $\mu$m. The sample chamber was created by laser drilling a hole of 300 $\mu$m in diameter. The cell was loaded with compressed He and closed at 1400 bar. The applied pressure was determined from ruby fluorescence. 
Room temperature data was collected in continuous $\phi$ rotation over 68$^\circ$ in shutterless readout of the EIGER2 X CdTe 9M detector (Dectris, Switzerland) in 0.25$^\circ$ steps. For the low temperature experiments diamond anvil cells made of CuNi$_2$Be were mounted in a Helium flow cryostat (ESRF) and data collected in continuous $\phi$ rotation over 64$^\circ$. For the ambient pressure experiment at low temperature the sample was mounted on a 300$\mu$m thick diamond plate and cooled with a small Helium flow cryostat (ESRF). Data was collected in continuous $\phi$ rotation over 70$^\circ$ in 0.25$^\circ$ steps. The detector distance was calibrated using CeO$_2$ powder to 185.5 mm for the room temperature data at ID27, 183.8 mm for low temperature at ID27 and 179.2 mm for low temperature at ID15B. The diffraction geometry was calibrated using a natural vanadinite single crystal and the flat-field of the detector was calibrated using air scattering at the employed x-ray energies. 
The key experimental results were obtained using monochromatic x-rays with a wavelength of $0.22290$ {\AA} for room temperature data, $0.3738$ {\AA} for low temperature at ID27 and $0.4099$ {\AA} for low temperature data at ID15B.

\begin{figure}[h]
	\begin{centering}
		\includegraphics[width=1\columnwidth]{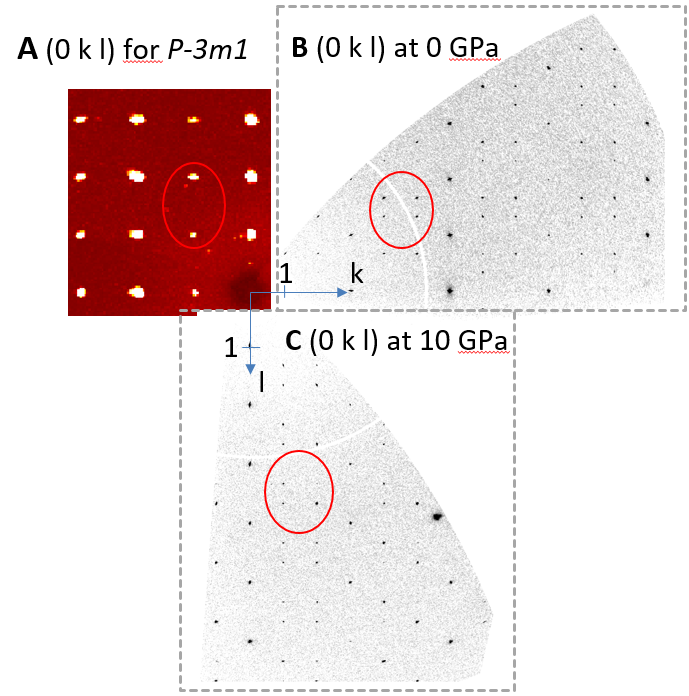} 
		\par\end{centering}
	\caption{Zonal diffraction maps. Zonal (0 k l) map sections of the single-crystal diffraction \textbf{a} in red for YCu$_3$(OH)$_{6}$Cl$_3$ with the $P\overline{3}m1$ subcell and two areas of single-crystal XRD of Y$_3$Cu$_9$(OH)$_{19}$Cl$_8$ with the supercell $R\overline{3}$ \textbf{a}  at 0 GPa and \textbf{b} 10 GPa. The sections are moved and scaled such that they match the indicated k, l-axis with a uniform origin. The changing reflexes moving from sub- to supercell are highlighted via a red circle.}
	\label{zonal}
\end{figure}

Y-Kapellasite crystallizes in the trigonal space group \#148 $R\overline{3}$ with $a=b=11.57$ \AA, $c=17,33$\AA~, which is a subgroup and supercell of YCu$_3$(OH)$_6$Cl$_3$  \cite{Kremer2025}. We display the zonal (0kl) maps of diffraction of the subcell in red in Fig. \ref{zonal}a, which clearly differ from the supercell (see Fig. \ref{zonal}b,c) independent of pressure. In the supercell, no partial occupancies exist and the Y atoms move out of the kagome plane, leading to a buckling of the kagome realized by two crystallographically inequivalent Cu positions displayed in Fig. 2. The buckling shortens one of the bonds to 3.26 \AA~(Cu1-Cu2 green line in Fig. 2) corresponding to the magnetic exchange path $J'$, while the others remain around 3.38 \AA. This squeezes the Cu-O-Cu bonding angle to a lower one around 110\degree, while the others remain at 116\degree~ and 117\degree~(see Fig. 2c). The magnetic substructure of the distorted kagome consists of the orange Cu1-Cu1 hexagons with the exchange interactions in orange $J_{\hexagon}$ linked by the blue $J$ from Cu2-Cu1 and the green $J'$ Cu1-Cu2 bonds.

In table \ref{tab:XRDres1} and \ref{tab:XRDres2} two representative refinement results are given for room temperature and 3 K.

\begin{table}[h]
    \centering
\begin{tabular}{ccccc}
{\footnotesize{}p (GPa)} & {\footnotesize{}a} & {\footnotesize{}c} & {\footnotesize{}R$_{1}$} & {\footnotesize{}wR$_{1}$}\tabularnewline
\hline 
{\footnotesize{}0} & {\footnotesize{}11.5691(5)} & {\footnotesize{}17.3305(10)} & {\footnotesize{}0.0351} & {\footnotesize{}0.1050}\tabularnewline
 &  &  &  & \tabularnewline
{\footnotesize{}atom} & {\footnotesize{}x} & {\footnotesize{}y} & {\footnotesize{}z} & {\footnotesize{}U$_{iso}$}\tabularnewline
\hline 
{\footnotesize{}Y1} & {\footnotesize{}0.000000} & {\footnotesize{}0.000000} & {\footnotesize{}0.37199(3)} & {\footnotesize{}0.01860(18)}\tabularnewline
{\footnotesize{}Y2} & {\footnotesize{}0.666667} & {\footnotesize{}0.333333} & {\footnotesize{}0.333333} & {\footnotesize{}0.0200(2)}\tabularnewline
{\footnotesize{}Cu1} & {\footnotesize{}0.32973(4)} & {\footnotesize{}0.17126(4)} & {\footnotesize{}0.33701(3)} & {\footnotesize{}0.01873(19)}\tabularnewline
{\footnotesize{}Cu2} & {\footnotesize{}0.166667} & {\footnotesize{}0.333333} & {\footnotesize{}0.333333} & {\footnotesize{}0.0191(2)}\tabularnewline
{\footnotesize{}Cl1} & {\footnotesize{}0.666667} & {\footnotesize{}0.333333} & {\footnotesize{}0.49507(10)} & {\footnotesize{}0.0276(4)}\tabularnewline
{\footnotesize{}Cl2} & {\footnotesize{}0.00414(10)} & {\footnotesize{}0.33576(10)} & {\footnotesize{}0.45096(8)} & {\footnotesize{}0.0277(3)}\tabularnewline
{\footnotesize{}O1} & {\footnotesize{}0.000000} & {\footnotesize{}0.000000} & {\footnotesize{}0.500000} & {\footnotesize{}0.058(3)}\tabularnewline
{\footnotesize{}O2} & {\footnotesize{}0.3357(2)} & {\footnotesize{}0.4933(2)} & {\footnotesize{}0.36848(18)} & {\footnotesize{}0.0187(5)}\tabularnewline
{\footnotesize{}H1} & {\footnotesize{}0.337264} & {\footnotesize{}0.508246} & {\footnotesize{}0.424228} & {\footnotesize{}0.028}\tabularnewline
{\footnotesize{}O3} & {\footnotesize{}0.4779(3)} & {\footnotesize{}0.3411(2)} & {\footnotesize{}0.37711(15)} & {\footnotesize{}0.0173(5)}\tabularnewline
{\footnotesize{}H2} & {\footnotesize{}0.475836} & {\footnotesize{}0.344910} & {\footnotesize{}0.433556} & {\footnotesize{}0.026}\tabularnewline
{\footnotesize{}O4} & {\footnotesize{}0.1964(2)} & {\footnotesize{}0.2018(3)} & {\footnotesize{}0.39012(15)} & {\footnotesize{}0.0181(5)}\tabularnewline
{\footnotesize{}H3} & {\footnotesize{}0.217421} & {\footnotesize{}0.224699} & {\footnotesize{}0.444724} & {\footnotesize{}0.027}\tabularnewline
 &  &  &  & \tabularnewline
{\footnotesize{}p (GPa)} & {\footnotesize{}a} & {\footnotesize{}c} & {\footnotesize{}R$_{1}$} & {\footnotesize{}wR$_{1}$}\tabularnewline
\hline 
{\footnotesize{}2.52} & {\footnotesize{}11.4351(4)} & {\footnotesize{}16.8341(3)} & {\footnotesize{}0.0486} & {\footnotesize{}0.1298}\tabularnewline
 &  &  &  & \tabularnewline
{\footnotesize{}atom} & {\footnotesize{}x} & {\footnotesize{}y} & {\footnotesize{}z} & {\footnotesize{}U$_{iso}$}\tabularnewline
\hline 
{\footnotesize{}Y1} & {\footnotesize{}0.666667} & {\footnotesize{}0.333333} & {\footnotesize{}0.70209(4)} & {\footnotesize{}0.0211(2)}\tabularnewline
{\footnotesize{}Y2} & {\footnotesize{}0.333333} & {\footnotesize{}0.666667} & {\footnotesize{}0.666667} & {\footnotesize{}0.0166(2)}\tabularnewline
{\footnotesize{}Cu1} & {\footnotesize{}0.50713(6)} & {\footnotesize{}0.50421(6)} & {\footnotesize{}0.67024(3)} & {\footnotesize{}0.0166(2)}\tabularnewline
{\footnotesize{}Cu2} & {\footnotesize{}0.833333} & {\footnotesize{}0.666667} & {\footnotesize{}0.666667} & {\footnotesize{}0.0178(2)}\tabularnewline
{\footnotesize{}Cl1} & {\footnotesize{}0.66498(12)} & {\footnotesize{}0.66247(12)} & {\footnotesize{}0.54784(5)} & {\footnotesize{}0.0226(3)}\tabularnewline
{\footnotesize{}Cl2} & {\footnotesize{}0.333333} & {\footnotesize{}0.666667} & {\footnotesize{}0.50255(9)} & {\footnotesize{}0.0251(4)}\tabularnewline
{\footnotesize{}O1} & {\footnotesize{}0.6716(3)} & {\footnotesize{}0.5337(4)} & {\footnotesize{}0.72421(16)} & {\footnotesize{}0.0169(6)}\tabularnewline
{\footnotesize{}H1} & {\footnotesize{}0.673371} & {\footnotesize{}0.554892} & {\footnotesize{}0.780748} & {\footnotesize{}0.025}\tabularnewline
{\footnotesize{}O2} & {\footnotesize{}0.5290(4)} & {\footnotesize{}0.6738(3)} & {\footnotesize{}0.71237(15)} & {\footnotesize{}0.0150(6)}\tabularnewline
{\footnotesize{}H2} & {\footnotesize{}0.534838} & {\footnotesize{}0.677473} & {\footnotesize{}0.770481} & {\footnotesize{}0.022}\tabularnewline
{\footnotesize{}O3} & {\footnotesize{}0.4910(3)} & {\footnotesize{}0.3354(3)} & {\footnotesize{}0.62907(18)} & {\footnotesize{}0.0163(7)}\tabularnewline
{\footnotesize{}H3} & {\footnotesize{}0.504035} & {\footnotesize{}0.337004} & {\footnotesize{}0.571469} & {\footnotesize{}0.024}\tabularnewline
{\footnotesize{}O4} & {\footnotesize{}0.666667} & {\footnotesize{}0.333333} & {\footnotesize{}0.833333} & {\footnotesize{}0.057(4)}\tabularnewline
 \end{tabular}
\caption{
    Single crystal refinement results of two representative room temperature diffraction results carried out under 0 and 2.52 GPa pressure. Note, that the two data sets are chosen in two different symmetry equivalent settings. Given are first a list of the pressure, temperature lattice constants and refinement values, followed by a list of the atomic positions}
    \label{tab:XRDres1}
\end{table}

\begin{table}[h]
    \centering
\begin{tabular}{ccccc}
{\footnotesize{}p (GPa)} & {\footnotesize{}a} & {\footnotesize{}c} & {\footnotesize{}R$_{1}$} & {\footnotesize{}wR$_{1}$}\tabularnewline
\hline 
{\footnotesize{}0} & {\footnotesize{}11.543(2)} & {\footnotesize{}17.154(3)} & {\footnotesize{}0.0449} & {\footnotesize{}0.0623}\tabularnewline
 &  &  &  & \tabularnewline
{\footnotesize{}atom} & {\footnotesize{}x} & {\footnotesize{}y} & {\footnotesize{}z} & {\footnotesize{}U$_{iso}$}\tabularnewline
\hline 
{\footnotesize{}Y1} & {\footnotesize{}1/3} & {\footnotesize{}2/3} & {\footnotesize{}2/3} & {\footnotesize{}0.00719(17)}\tabularnewline
{\footnotesize{}Y2} & {\footnotesize{}2/3} & {\footnotesize{}1/3} & {\footnotesize{}0.70473(3)} & {\footnotesize{}0.00835(14)}\tabularnewline
{\footnotesize{}Cu1} & {\footnotesize{}0.49533(4)} & {\footnotesize{}0.49150(4)} & {\footnotesize{}0.66976(2)} & {\footnotesize{}0.00718(17)}\tabularnewline
{\footnotesize{}Cu2} & {\footnotesize{}5/6} & {\footnotesize{}2/3} & {\footnotesize{}2/3} & {\footnotesize{}0.0076(2)}\tabularnewline
{\footnotesize{}Cl1} & {\footnotesize{}0.33609(8)} & {\footnotesize{}0.33107(8)} & {\footnotesize{}0.54935(5)} & {\footnotesize{}0.0097(3)}\tabularnewline
{\footnotesize{}Cl2} & {\footnotesize{}1/3} & {\footnotesize{}2/3} & {\footnotesize{}0.50522(7)} & {\footnotesize{}0.0088(3)}\tabularnewline
{\footnotesize{}O1} & {\footnotesize{}0.4651(3)} & {\footnotesize{}0.3279(2)} & {\footnotesize{}0.72404(15)} & {\footnotesize{}0.0098(9)}\tabularnewline
{\footnotesize{}H1} & {\footnotesize{}0.435436} & {\footnotesize{}0.32471} & {\footnotesize{}0.777934} & {\footnotesize{}0.0117}\tabularnewline
{\footnotesize{}O2} & {\footnotesize{}0.5228(3)} & {\footnotesize{}0.6588(2)} & {\footnotesize{}0.62334(14)} & {\footnotesize{}0.0084(9)}\tabularnewline
{\footnotesize{}H2} & {\footnotesize{}0.540177} & {\footnotesize{}0.654953} & {\footnotesize{}0.567746} & {\footnotesize{}0.0101}\tabularnewline
{\footnotesize{}O3} & {\footnotesize{}0.4911(2)} & {\footnotesize{}0.3310(2)} & {\footnotesize{}1.03524(15)} & {\footnotesize{}0.0091(9)}\tabularnewline
{\footnotesize{}H3} & {\footnotesize{}0.490409} & {\footnotesize{}0.330699} & {\footnotesize{}1.092374} & {\footnotesize{}0.01}\tabularnewline
{\footnotesize{}O4} & {\footnotesize{}2/3} & {\footnotesize{}1/3} & {\footnotesize{}5/6} & {\footnotesize{}0.044(3)}\tabularnewline
 &  &  &  & \tabularnewline

{\footnotesize{}p (GPa)} & {\footnotesize{}a} & {\footnotesize{}c} & {\footnotesize{}R$_{1}$} & {\footnotesize{}wR$_{1}$}\tabularnewline
\hline 
{\footnotesize{}2.6} & {\footnotesize{}11.4270(4)} & {\footnotesize{}16.810(4)} & {\footnotesize{}0.0488} & {\footnotesize{}0.0638}\tabularnewline
 &  &  &  & \tabularnewline
{\footnotesize{}atom} & {\footnotesize{}x} & {\footnotesize{}y} & {\footnotesize{}z} & {\footnotesize{}U$_{iso}$}\tabularnewline
\hline 
{\footnotesize{}Y1} & {\footnotesize{}2/3} & {\footnotesize{}1/3} & {\footnotesize{}1/3} & {\footnotesize{}0.0189(19)}\tabularnewline
{\footnotesize{}Y2} & {\footnotesize{}1/3} & {\footnotesize{}2/3} & {\footnotesize{}0.30186(16)} & {\footnotesize{}0.0371(16)}\tabularnewline
{\footnotesize{}Cu1} & {\footnotesize{}2/3} & {\footnotesize{}5/6} & {\footnotesize{}1/3} & {\footnotesize{}0.0085(2)}\tabularnewline
{\footnotesize{}Cu2} & {\footnotesize{}0.49339(5)} & {\footnotesize{}0.49617(5)} & {\footnotesize{}0.33044(10)} & {\footnotesize{}0.0089(2)}\tabularnewline
{\footnotesize{}Cl1} & {\footnotesize{}1/3} & {\footnotesize{}2/3} & {\footnotesize{}0.5009(4)} & {\footnotesize{}0.0151(4)}\tabularnewline
{\footnotesize{}Cl2} & {\footnotesize{}0.33474(10)} & {\footnotesize{}0.33736(10)} & {\footnotesize{}0.4525(3)} & {\footnotesize{}0.0133(3)}\tabularnewline
{\footnotesize{}O1} & {\footnotesize{}0.3286(3)} & {\footnotesize{}0.4675(3)} & {\footnotesize{}0.2757(6)} & {\footnotesize{}0.0131(7)}\tabularnewline
{\footnotesize{}H1} & {\footnotesize{}0.326227} & {\footnotesize{}0.444822} & {\footnotesize{}0.229196} & {\footnotesize{}0.0157}\tabularnewline
{\footnotesize{}O2} & {\footnotesize{}0.5101(3)} & {\footnotesize{}0.6646(3)} & {\footnotesize{}0.3712(7)} & {\footnotesize{}0.0138(7)}\tabularnewline
{\footnotesize{}H2} & {\footnotesize{}0.513137} & {\footnotesize{}0.664169} & {\footnotesize{}0.419943} & {\footnotesize{}0.0166}\tabularnewline
{\footnotesize{}O3} & {\footnotesize{}0.6599(3)} & {\footnotesize{}0.5218(3)} & {\footnotesize{}0.3788(6)} & {\footnotesize{}0.0124(6)}\tabularnewline
{\footnotesize{}H3} & {\footnotesize{}0.656984} & {\footnotesize{}0.535464} & {\footnotesize{}0.426418} & {\footnotesize{}0.0149}\tabularnewline
{\footnotesize{}O4} & {\footnotesize{}1/3} & {\footnotesize{}2/3} & {\footnotesize{}0.166667} & {\footnotesize{}0.048(4)}\tabularnewline
\end{tabular}
\caption{
    Single crystal refinement results of two representative low temperature diffraction results carried out at 3 K under 0 and 2.6 GPa pressure. Note, that the two data sets are chosen in two different symmetry equivalent settings. Given are first a list of the pressure, temperature lattice constants and refinement values, followed by a list of the atomic positions}
    \label{tab:XRDres2}
\end{table}

\newpage
\thispagestyle{empty}
\mbox{}
\newpage

\subsection*{Exchange interaction}
\begin{figure}[h]
	\begin{centering}
		\includegraphics[width=1\columnwidth]{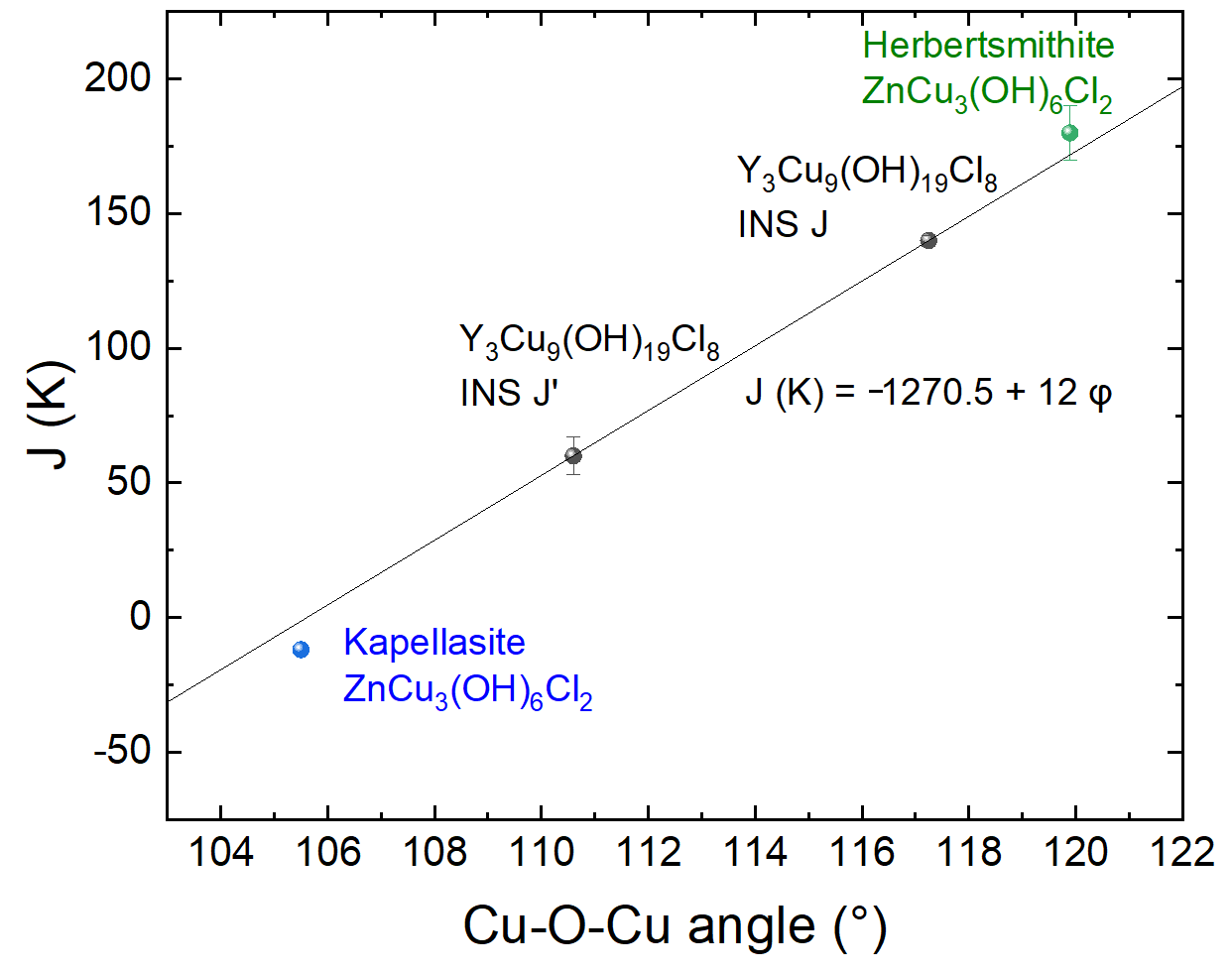}
		\par\end{centering}
	\caption{\textbf{Angle dependence of the $J$.} Superexchange interaction $J$ in Kelvin versus the copper-oxide-copper angle for Y-Kapellasite (black) \cite{Chatterjee2023} kapellasite (blue) \cite{Kermarrec2014}and herbertsmithite (green) \cite{Khuntia2020}.}
	\label{superexch}
\end{figure}

In our compound we are dealing with a hydroxychloride-system, where a crossover for FM (here negative defined) to AFM (here positive) is expected around 105$\degree$ \cite{Boldrin2015}. For our exact case, we have two anchorpoints from our inelastic neutron scattering experiment, where we find that $J_{\hexagon} \approx J = 140$ K and $J'=63\pm7$ K determined at a temperature of 1.55 K \cite{Chatterjee2023}. Using our lowest temperature diffraction result summarized in Fig. 2 we know the angles of $J_{\hexagon}$ and $J$ are 117 $\degree$ and 117.5 $\degree$ averaging to $117.25 \pm0.25 \degree$ and $J'$ has the angle $110.6\degree$ (see Fig. \ref{superexch}. Consequently we can use a linear fit to work out the change of our exchange interactions for the angles that do vary within 110 to 118$\degree$ as we observe in our XRD analysis. On top we have determined in previous works for the closely related materials kapellasite ZnCu$_3$(OH)$_6$Cl$_2$ an exchange interaction of -12 K \cite{Kermarrec2014} with only one Cu-O-Cu bond of an angle of 105.5$\degree$ and herbertsmithite ZnCu$_3$(OH)$_6$Cl$_2$ an exchange interaction of 180 K \cite{Khuntia2020} and kagome bond angle of 118.9$\degree$ \cite{Kremer2025}. These additional points of related systems support the range and estimation of our exchange interactions within errorbars.

\subsection*{Comparison of external and chemical pressure}

While external pressure provides a clean control parameter that compresses the lattice and drives the system toward increased isotropy, we also investigate the effects of chemical pressure \cite{Krieger2025}. Chemical pressure is realized through substitution of the $A$-site ion, which is typically a rare-earth element. With the exception of Y and Lu (La being too large to stabilize the structure), this substitution introduces a magnetic ion at the center of the kagome hexagon. However, the rare-earth magnetism plays a significant role only at high magnetic fields, as it couples to the Cu moments only indirectly \cite{Krieger2025}.

\begin{figure}[h]
	\begin{centering}
		\includegraphics[width=1\columnwidth]{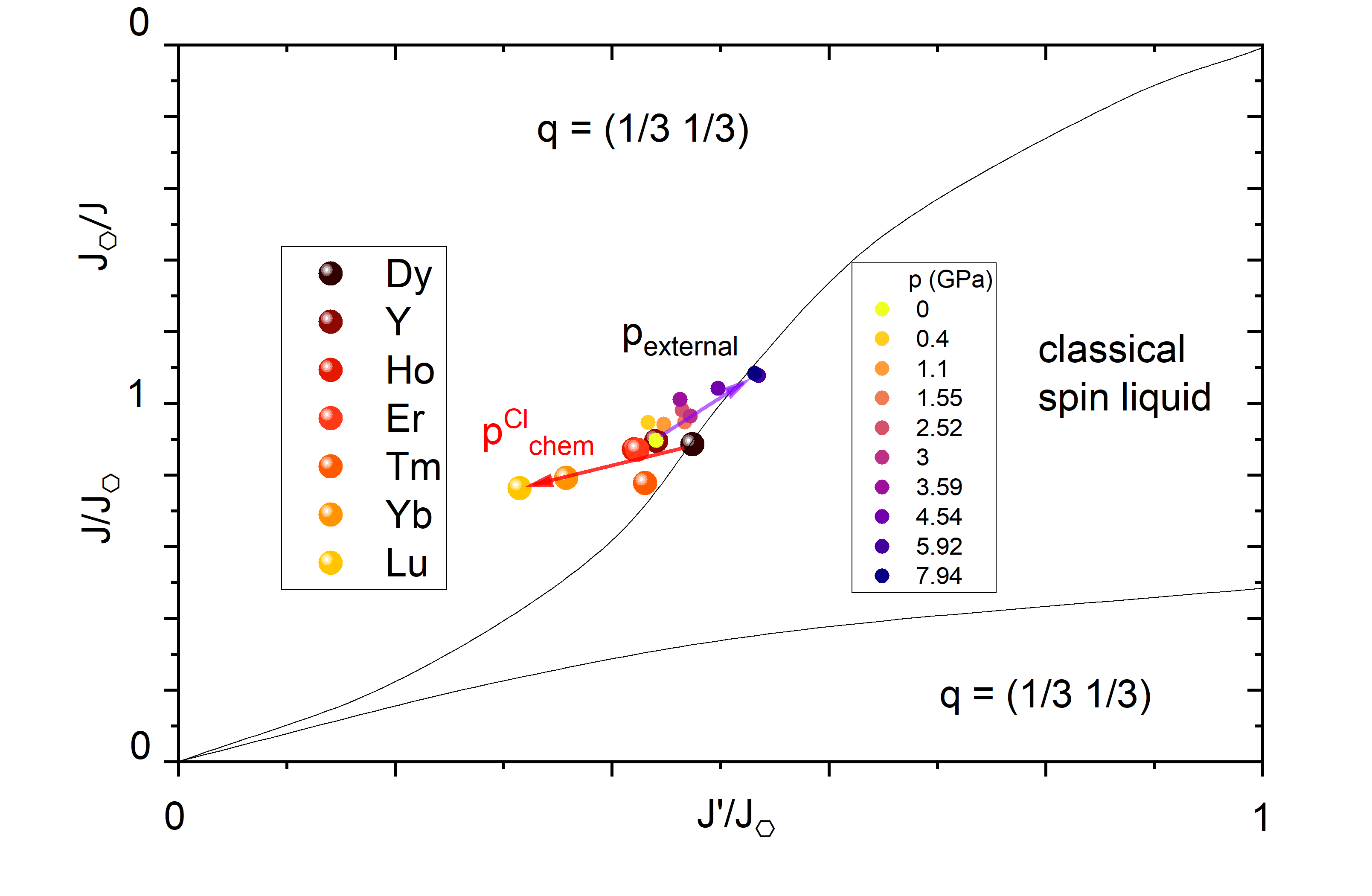}
		\par\end{centering}
	\caption{Classical phase diagram of the distorted kagome antiferromagnet as a function of the exchange couplings $J_{\hexagon}$, $J$, and $J'$ \cite{Hering2022}. The left (right) region corresponds to an antiferromagnetic in-plane $(1/3,1/3)$ ordered phase (spin-liquid, SL). Estimated exchange parameters obtained from high-temperature data are shown for chemical pressure (brown to yellow symbols) and external pressure (yellow to blue symbols), with pressure values color-coded as indicated in the legend \cite{Krieger2025}.}
	\label{chem}
\end{figure}

Figure~\ref{chem} compares the evolution of the system under external and chemical pressure in the classical phase diagram. Arrows indicate the direction of increasing pressure. In contrast to external pressure, chemical pressure allows for both positive and negative tuning: larger $A$-site ions relative to Y effectively apply negative pressure, with Y serving as a common reference point. Although both tuning parameters lead to an overall lattice contraction, they modify the local geometry in qualitatively different ways. Chemical pressure, induced by a smaller $A$-site ion at the center of the hexagon, locally pulls on the surrounding oxygen atoms and bends the Cu–O–Cu bonds. External pressure, by contrast, pushes the Y ions back into the hexagon, flattening the Cu–O–Cu bonds. This distinction highlights external pressure as a particularly effective tuning parameter for controlling the exchange anisotropy.

\subsection*{Optics}
Pressure-dependent optical studies were carried out at SMIS beamline of Synchrotron Soleil, France. 
The vector of electric field was parallel with the basal plane of the unit cell, which means that only the E$_{\mathrm{u}}$ phonon modes are observable \cite{Dolezal2024}. The optical measurements in diamond anvil cells limit the accessible signal and ranges, which are not hampered by the absorption of the diamond windows. Due to the samples thickness in case of strong absorption we are losing the shape of the maxima, visible in comparison to the dotted line in Fig. 4 of the main text.  The spectral range below 150~cm$^{-1}$ was not detectable due to strong water vapor absorption in the far-infrared. In this range, our ambient-pressure measurements have evidenced magneto-elastic coupling \cite{Dolezal2024} leading to two subtle H-based structural transition below 33 K \cite{Chatterjee2023} unresolvable by XRD.

At SMIS we used a membrane driven DAC of the Le Toullec type, with Ilas diamond anvils and 500/600 $\mu$m culet, stainless steel gaskets preindented to a final thickness of about 50 $\mu$m, where a hole has been drilled by electric discharge machine. The pressure transmitting medium was Polyethylene (PE) and NaCl in the FIR and MIR ranges, respectively \cite{Celeste2019}.
 The pressure cell was mounted in a liquid He flow cryostat equipped with BaF$_2$ windows for the MIR range and polypropilene ones for the FIR range. The spectra were collected with a ThermoFisher iS50 FTIR spectrometer, with synchrotron radiation as a light source, KBr (solid substrate) beamsplitter and a MCT (liquid He-cooled Si bolometer) detector for the MIR (FIR) range. The spectrometer was coupled to an in-house developed horizontal IR microscope having custom reflective Schwarzschild objectives with large working distance.

\begin{figure}[h]
	\begin{centering}
		\includegraphics[width=1\columnwidth]{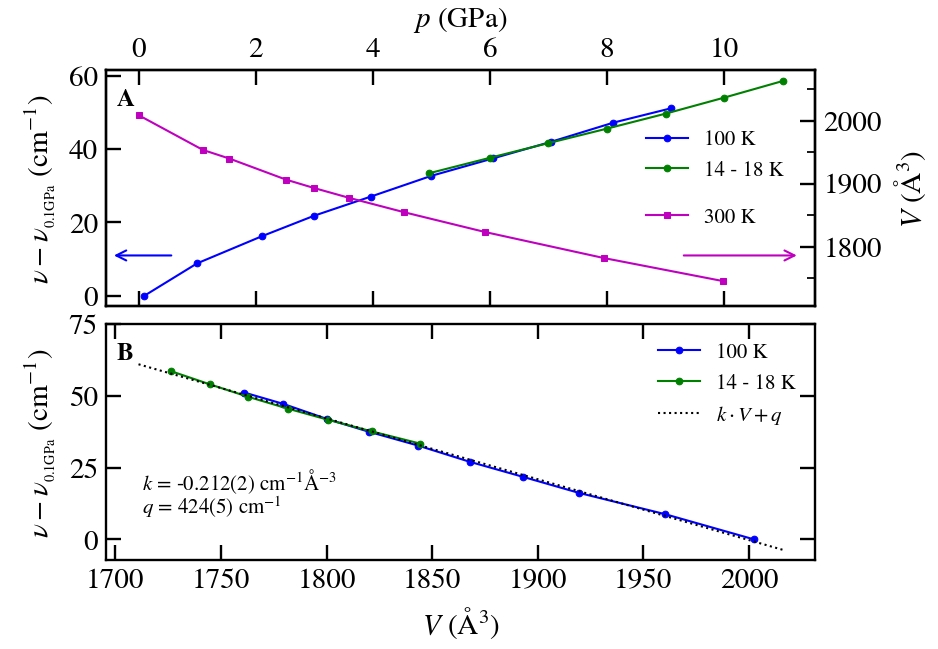}
		\par\end{centering}
	\caption{\textbf{Phonon shift.} \textbf{a} Pressure dependence of unit cell volume and shift of phonon modes in spectral region2, results of the fit (For details see the main text.). \textbf{b} The volume dependence of phonon mode shift constructed from the above plot. The dependence is fitted by a linear function.}
	\label{Gruneisen_parameter}
\end{figure}

We estimate the Grüneisen parameter $\gamma$ Ref.~\cite{Gruneisen_1912}, $\gamma = - \frac{V}{\nu} \frac{d \nu}{dV}$. The details and results of such fits are shown in Fig. \ref{Gruneisen_parameter}. The dependence $\nu - \nu_{0.1 \mathrm{GPa}} = a$ together with the unit cell volume is shown in Fig.~S2 (a). The $\nu(V)$ dependence is then shown in panel (b), which leads to a linear dependence proving the direct link of Cu O modes to the unit cell volume. With the average frequency of 459 cm$^{-1}$ and volume of 2150 \r{A}$^{3}$ we find $\gamma = 0.85$. Since we have two Cu and three O Wyckoff positions we have five different Cu-O bonds. Thus individual modes in region 2 will have slightly different pressure dependence of their positions as can be seen from Fig.~S3 and S4, but their blue shift is quite robust. Summarized

\begin{figure}[t]
	\begin{centering}
		\includegraphics[width=1\columnwidth]{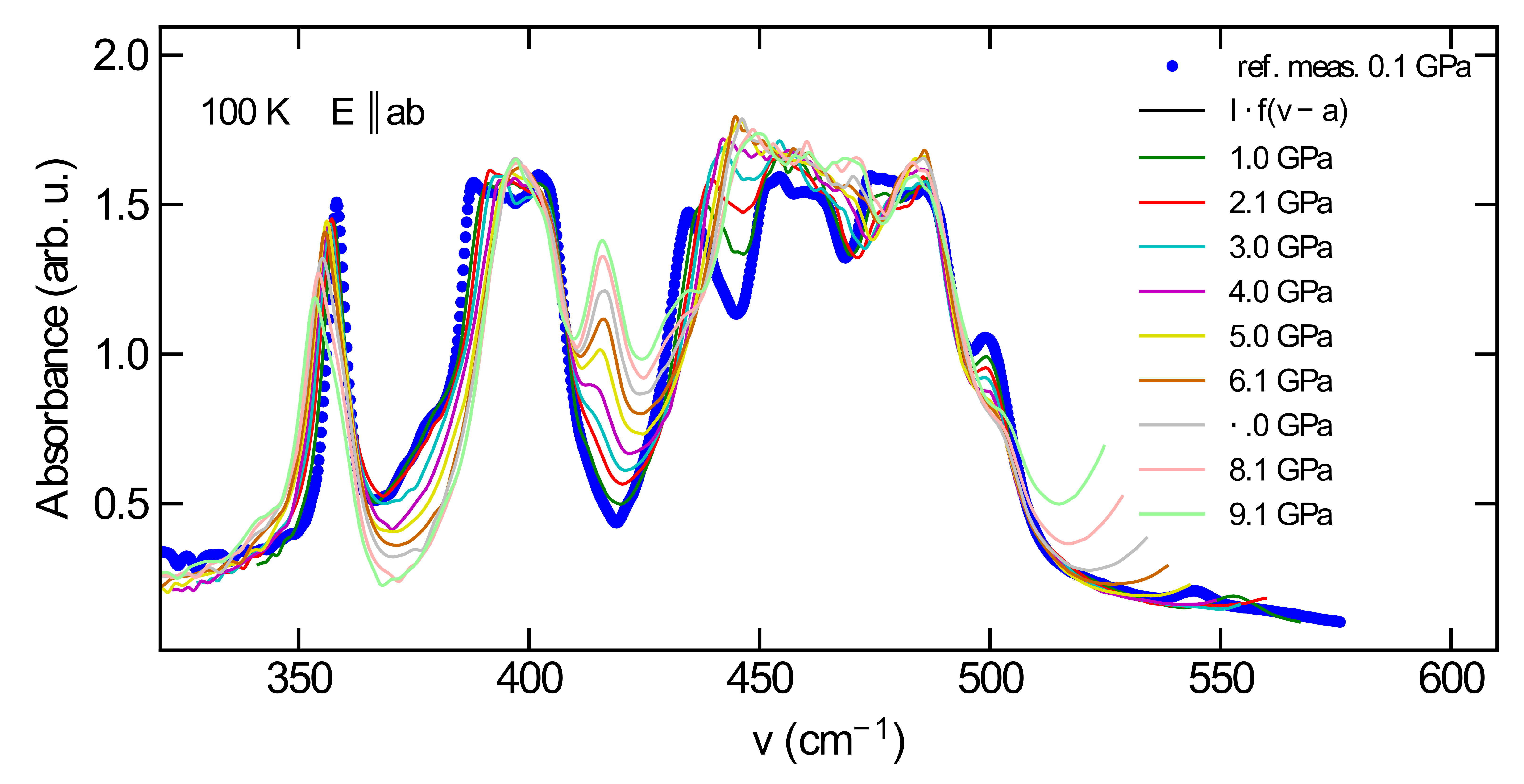}
		\par\end{centering}
	\caption{\textbf{High temperature low wavenumber region of the absorbance.} Results of the shifted spectral intensity $I f(\nu - a)$ in region 2 at 100~K. As the reference the spectral intensity at 0.1~GPa was used. For details see the main text.}
	\label{Shift_of_pressure_data_100K}
\end{figure}

\begin{figure}[t]
	\begin{centering}
		\includegraphics[width=1\columnwidth]{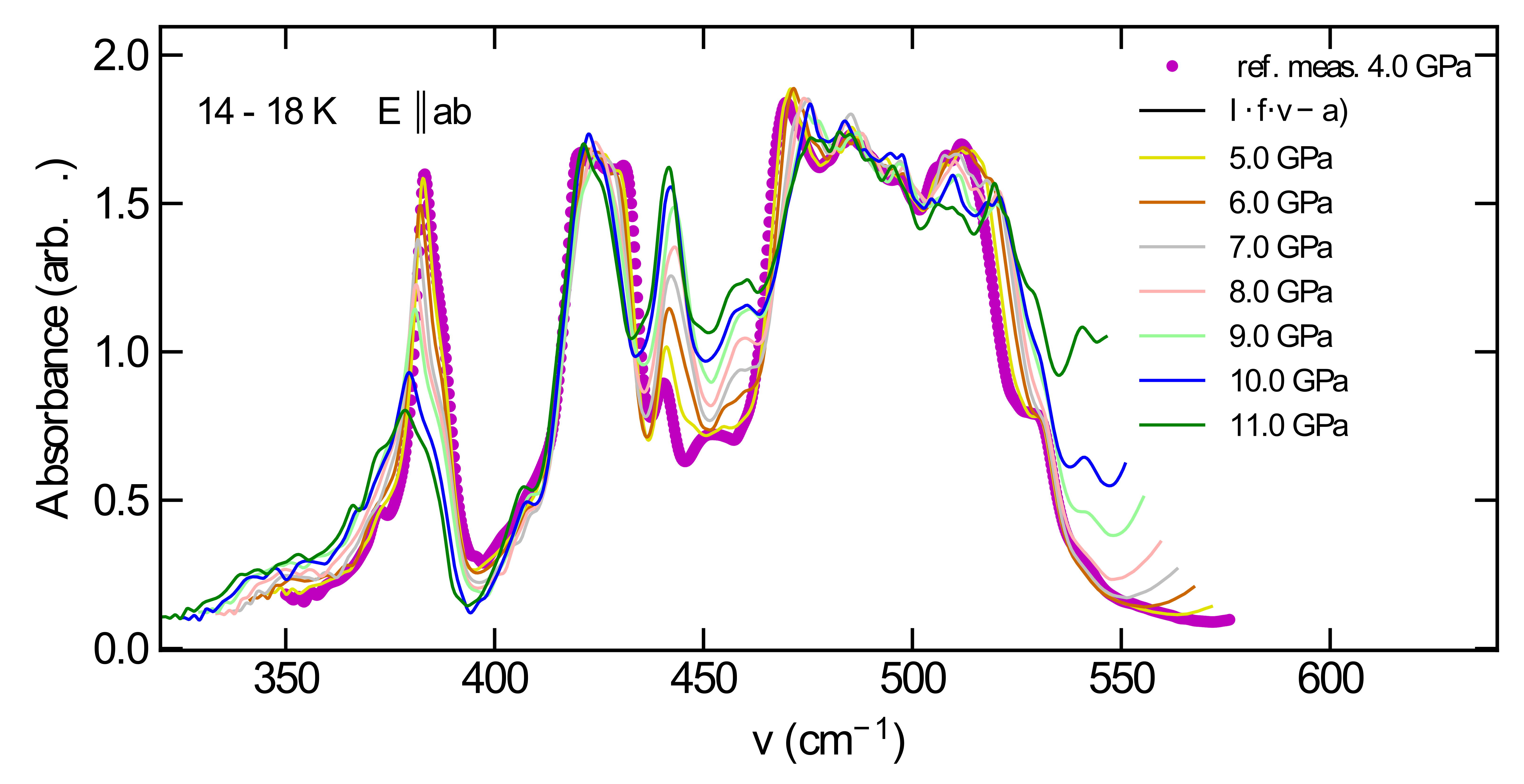}
		\par\end{centering}
	\caption{\textbf{Low temperature low wavenumber region of the absorbance. } Results of the shifted spectral intensity $I f(\nu - a)$ in region 2 at 14-18~K. As the reference the spectral intensity at 4.0~GPa was used. For details see the main text.}
	\label{Shift_of_pressure_data_14_18K}
\end{figure}

 The pressure dependence of the mode frequencies $\nu$ linked to the volume is captured by the Grüneisen parameter $\gamma$ Ref.~\cite{Gruneisen_1912} $\gamma = - \frac{V}{\nu} \frac{d \nu}{dV}$. To subtract the $\nu$ dependence of the spectral region 2 on the pressure, we took the lowest pressure data as the reference and the other pressure spectra were shifted and scaled to this reference data via $I f(\nu - a)$. $I$ and $a$ are free parameters for the fit. The results of such fits are shown in Fig. \ref{Shift_of_pressure_data_100K} and Fig. \ref{Shift_of_pressure_data_14_18K}. We show for a selected frequency range the change of spectral intensity with increasing pressure at 100 K and 14-18 K.
 As we have two Cu and three O Wyckoff positions we have five different Cu-O bonds  individual modes will have slightly different pressure dependence of their positions as can be seen from Fig.s~\ref{Shift_of_pressure_data_100K} and \ref{Shift_of_pressure_data_14_18K}, but their blue shift is robust.

\maketitle
\bibliography{library.bib}